\documentclass[prl,aps,twocolumn,superscriptaddress,longbibliography,nofootinbib,floatfix]{revtex4-1}
\usepackage[utf8]{inputenc}
\usepackage{multirow}
\usepackage{amsmath}
\usepackage{amsfonts}
\usepackage{graphicx}
\usepackage{array}
\usepackage{hyperref}
\usepackage{braket}
\usepackage{xcolor}
\usepackage{verbatim}
\usepackage{subfig}
\usepackage{ulem}
\usepackage{physics}

\usepackage{amssymb}

\newcommand{\threej}[6]{ \begin{pmatrix}
  #1 & #2 & #3 \\
  #4 & #5 & #6 
 \end{pmatrix}}

\newcommand{\sixj}[6]{ \begin{Bmatrix}
  #1 & #2 & #3 \\
  #4 & #5 & #6 
 \end{Bmatrix}}

\DeclareFontFamily{U}{mathb}{\hyphenchar\font45}
\DeclareFontShape{U}{mathb}{m}{n}{<5> <6> <7> <8> <9> <10> gen * mathb
<10.95> mathb10 <12> <14.4> <17.28> <20.74> <24.88> mathb12}{}
\DeclareSymbolFont{mathb}{U}{mathb}{m}{n}
\DeclareMathSymbol{\rcirclearrow}{0}{mathb}{'367}

\begin{document}

\title{Optical reconstruction of collective density matrix of qutrit}
\date{\today}

\author{Marek Kopciuch}
\affiliation{Institute of Physics, Jagiellonian University in Krak\'ow, {\L}ojasiewicz 11, 30-348 Krak\'ow, Poland}

\author{Szymon Pustelny}
\affiliation{Institute of Physics, Jagiellonian University in Krak\'ow, {\L}ojasiewicz 11, 30-348 Krak\'ow, Poland}

\begin{abstract}
    Reliable tomography of a quantum state of atoms in room-temperature vapors offers interesting applications in quantum-information science. To step toward the applications, here, we theoretical investigate a technique of reconstruction of a collective density matrix of atoms in a state of a total angular momentum equal to unity. By studying transformation of density matrix due to magnetic-field pulses and deriving explicit relations between optical signals and specific density-matrix elements, we demonstrate full reconstruction of quantum states. Numerical simulations allow us to demonstrate that the reconstruction fidelity of pure and mixed states may exceed 0.95 and that the technique is robust against noise/uncertainties.
\end{abstract}
\maketitle

\section{Introduction}
One of the most important challenges of quantum-information science is the reconstruction of an unknown quantum state \cite{Paris2004Quantum, Liu2019PulsedSystems, Mancini1997Quantum, Briegel2009Measurement-basedComputation, Cirac2004QuantumOptics}.  The reconstruction requires measurements of different non-commuting observables on a set of systems occupying the same quantum state.  Unfortunately, due to the no-cloning theorem \cite{Wootters1982ACloned}, preparation of identical copies of a single quantum state is impossible.  Utilization of a macroscopic system, composed of a large number of particles and described by a collective quantum state, enables measurements over many elements at the same time \cite{Cirac2004QuantumOptics, Deutsch2010Quantum}. Interpretation of such a reconstruction is straightforward in media composed of particles in a (statistically) identical quantum state. A specific example of such a system is the Bose-Einstein condensate \cite{Mancini1997Quantum}. The situation is different in room-temperature atomic vapors, where inhomogeneous transition broadening leads to a velocity-dependent interaction with external fields (the Doppler effect) and, as a result, atoms in various quantum states are simultaneously present in the medium.  Nevertheless, even in such a case, there are distinct examples in which atoms may be considered to be statistically occupying the same quantum state.  A specific example of this situation is an atomic vapor contained in a paraffin-coated cell, where atoms freely travel between the container walls and their polarization is preserved in wall collisions \cite{Zhivun2016VectorCells, Graf2005RelaxationCells, Balabas2010PolarizedTime}.  Thus, the interaction of atoms is averaged over a whole cell volume \cite{Zhivun2016VectorCells, Graf2005RelaxationCells}).  Although in reality, the whole-system state is a statistical average over quantum states of its microscopic elements, the system still possesses many features of a quantum state, revealing numerous quantum properties (shot noise \cite{Lucivero2016Squeezed-lightSpectroscopy, Lucivero2014Shot-noise-limitedTemperature}, squeezing \cite{Lucivero2016Squeezed-lightSpectroscopy}, entanglement \cite{Julsgaard2001ExperimentalObjects}, etc.). In turn, it is interesting from the point of view of quantum-state tomography. 

In this work, we present an approach to reconstruct the collective density matrix \cite{Sakurai2017ModernMechanics, Smith2004Continuous} of atoms in an atomic ensemble based on continuous weak measurements \cite{Deutsch2010Quantum, Smith2004Continuous, Pustelny2006Pump-probeLight}.  In measurements, a system under investigation is weakly perturbed during probing; hence, the attainable information is limited.  This fact can be regained by performing measurements at many microscopic subsystems (e.g., atoms) at the same time.  The approach was already implemented in quantum-state tomography of cold atoms (see, for example, Ref.~\cite{Deutsch2010Quantum}), however, contrary to previous works, our approach is described in a semiclassical picture, in which incident light is purely classical. 

The semiclassical approximation is a sufficient and conventional way to describe light-atom interaction when light is in a coherent state (e.g., laser light). This formalism is widely used to describe room-temperature and hot atomic gases \cite{Auzinsh2010OpticallyInteractions}, but can also be used to describe cold-atom experiments \cite{Auzinsh2010OpticallyInteractions, Sycz2018Atomic_stateExperiments}. Previous approaches achieved nondestructive measurements using far-detuned light, which allows one to use an effective ground-state Hamiltonian \cite{Deutsch2010Quantum}, neglecting the tensor part of the light-atom interaction. In our approach, we take into account entire light-atom interaction, specifically including the tensor contribution to the interaction, and calculate the evolution of the system using the first-order perturbation theory. As a result, we explore scenarios in which weak, yet arbitrary, detuning light is used. This is of particular importance in the case of Doppler-broadened media, where one needs to account for atoms with different velocity classes. Moreover, as shown below, by carefully choosing the detuning, one can gain additional information about the system from a single measurement.
Our approach offers the ability to derive analytical formulas for atomic evolution rather than numerical simulation of the evolution, as previously done \cite{Smith2004Continuous}. Because the overall evolution of such a system is complicated, we perform a series of approximations, suitable for alkali vapors at room temperature. It allows us to relate polarization rotation and ellipticity change with the quantum properties of atoms (density-matrix elements). In turn, this enables the reconstruction of an arbitrary quantum-mechanical state of atoms based on the observed signals.  Additionally, we demonstrate application of the approach for the full reconstruction of a density matrix of a three-level system (qutrit) formed within magnetic sublevels of long-lived ground state \cite{Budker2005Microwave,Balabas2010PolarizedTime}. The technique is implemented for the reconstruction of both pure and mixed states, and its precision and robustness against different limitations (noise and uncertainties) are investigated.

\section{Relation between observables and quantum-state properties}
Optical tomography of a quantum state is based on measurements of specific properties of light propagating through a medium consisting of atoms whose quantum state one wants to reconstruct.  Particular examples of observables that provide such information are: the polarization angle $\alpha$ and the degree of ellipticity $\epsilon$.  In the appendix A, the relations of these two observables with the density matrix elements characterizing the atoms of the ground state $f$ and the excited state $F$, where $f$ and $F$ denote the total angular momentum in the ground- and excited state, respectively, are derived. Although it is possible to derive the relation for any light polarization, here we limit ourselves to the case of $y$-polarized light propagating along the $z$-axis.  This gives us the relations for the polarization rotation $\partial_z\alpha$ and the ellipticity $\partial_z\epsilon$
\begin{equation}
    \partial_{z}\epsilon - i \partial_{z}\alpha = -\dfrac{\omega}{2 \varepsilon_0 E_0 c} U^{T}_{xq} \mathcal{P}^{q},
    \label{eq:RotationVsCoherences}
\end{equation}
where $U^{T}_{xq}$ is the transformation from the spherical to the Cartesian basis \cite{Auzinsh2010OpticallyInteractions} and $\mathcal{P}^{q}$ is the $q^\text{th}$ component of the complex polarization of the atomic vapor in the spherical basis
\begin{widetext}
    \begin{equation}
        \mathcal{P}^{q} = 2 N_{at} (-1)^{j+f+I+F+1} \sqrt{(2f+1)(2F+1)} \expval{j \norm{d^{(J)}} J} \sixj{j}{f}{I}{F}{J}{1} \left[ \sum_{m,\mu} (-1)^{m} \threej{f}{1}{F}{-m}{q}{\mu} \tilde{\rho}_{\mu m}(t) \right],
    \end{equation}
\end{widetext}
 where $\tilde{\rho}_{\mu m}$ is the optical-coherence amplitude between the ground state $m$ and the excited state $\mu$, $\tilde\rho_{\mu m}$ is given in the rotating wave approximation, $\omega$ is the frequency and $E_0$ is the amplitude of light, and $N_{at}$ denotes the atomic number density. $f$ and $F$ are the total angular momenta of the ground and excited state, respectively, $j$ and $J$ are the corresponding total electronic angular momenta, and $I$ is the nuclear spin. The reduced dipole matrix element $\expval{j  \norm{\widehat{\mathbf{d}}^{(J)}} J }$ is calculated in the $J$ manifold It should be noted that the angular-momentum composition of the system is given by the Wigner 3j and 6j symbols. Precise knowledge of the $\omega / \left(2 \varepsilon_0 E_0 c \right)$ factor, appearing in Eq.~\eqref{eq:RotationVsCoherences},  is of crucial importance for quantum-state tomography.

The evolution of optical coherence of atoms subjected to the magnetic field $B_z$ and interacting with weak $x$-polarized light (we only account for linear terms in $E_0$ and expand the density matrix as $\tilde{\rho} = \sum_{n} E_{0}^n\tilde{\rho}^{(n)} \approx \tilde{\rho}^{(0)} + E_0 \tilde{\rho}^{(1)}$), detuned from the optical transition by $\Delta$, is derived in appendix A.  For the current discussion, we simplify the relations by noting that (a) the ground-state relaxation rate $\gamma$ is typically several orders of magnitude smaller than the excited-state relaxation rate $\Gamma$, $\Gamma\gg\gamma$, and (a) the Larmor frequency $\Omega_L$ is small compared to the detuning $\Delta$, $\Delta\gg\Omega_L$.  This allows us to simplify the equation for the time evolution of the optical-coherence amplitude 
\begin{widetext}
    \begin{equation}
        \dot{\tilde{\rho}}_{\mu m}(t) \approx - \left( i \Delta + \dfrac{\Gamma}{2} \right) \tilde{\rho}_{\mu m}(t) + \dfrac{i \Omega_{R}}{2}(-1)^{j-f+3F+I-\mu} \sqrt{(2f+1)(2F+1)} \sixj{j}{f}{I}{F}{J}{1} \left[ \sum_{q,n} U_{yq}^{T} \threej{F}{1}{f}{-\mu}{q}{n} \rho_{nm}^{(0)}(t) \right]
        \label{eq:CoherenceVsTime}
    \end{equation}
\end{widetext}
where $\Omega_R = \expval{j  \norm{\widehat{\mathbf{d}}^{(J)}} J } E_0$ is the Rabi frequency, $\rho_{mn}^{(0)}$ is the zeroth order Zeeman coherence between the ground-state magnetic sublevels $m$ and $n$ (if $m=n$, $\rho_{mm}^{(0)}$ is the ground-state population of the sublevel $m$). It is noteworthy that we dropped the tilde above ground-state density-matrix elements, as RWA only applies to optical coherences and does not affect the ground-state density-matrix elements. It is important to note that because we are interested in the effects linear in $E_0$, Eq.~\eqref{eq:CoherenceVsTime} is independent of the excited-state evolution and depends only on the evolution of the $0^\text{th}$ order of the density matrix. The $\rho_{nm}^{(0)}(t)$ term describes the evolution of the ground-state manifold within uniform static magnetic field, assuming negligible back action (this is a good approximation of the ground-state density matrix). Using the Liouville equation, one can get the time-dependent formulas
\begin{subequations}
    \begin{eqnarray}
        \rho_{mm}^{(0)}(t) &=& \left(\rho_{nn}(0) -  \dfrac{1}{2f+1}\right) e^{-\gamma t} + \dfrac{1}{2f+1},\\
        \rho_{nm}^{(0)}(t) &=&\rho_{nm}(0)e^{-\gamma t} e^{- i (n-m) \Omega_{L} t},
    \end{eqnarray}
    \label{eq:GroundStateVsTime}
\end{subequations}

Taking a time derivative of Eq.~\eqref{eq:RotationVsCoherences} and combining it with Eqs.~\eqref{eq:CoherenceVsTime} and \eqref{eq:GroundStateVsTime} allows us to relate the time-dependent polarization rotation and ellipticity change with the density-matrix elements
\begin{widetext}
    \begin{equation}
        \dfrac{d}{dt}\left[ \partial_z \epsilon  - i \partial_z \alpha \right] = - \left( i \Delta + \dfrac{\Gamma}{2} \right) \left[ \partial_z \epsilon  - i \partial_z \alpha \right] + \dfrac{i \chi}{2} e^{-\gamma t} \left[ \expval{\hat{\alpha}_{R}} \sin (2 \Omega_L t) + \expval{\hat{\alpha}_{I}} \cos (2 \Omega_L t) - i \expval{\hat{\beta}} \right],
    \label{eq:RotationVsKappa}
    \end{equation}
\end{widetext}
where
\begin{equation}
\small
    \chi = \dfrac{N_{at}\omega \abs{\expval{j  \norm{\widehat{\mathbf{d}}^{(J)}} J }}^{2} }{2 \varepsilon_0 c \hbar} (-1)^{2j+2J}(2f+1)(2F+1) \sixj{j}{f}{I}{F}{J}{1}^2\,,
\end{equation}
and $\hbar$ is the reduced Planck constant. Here the density-matrix elements are grouped into expectation values of the effective observables
\begin{widetext}
    \begin{subequations}
        \begin{eqnarray}
            \hat{\alpha}_{R} &=& \sum_{m=-f}^{f-2} \threej{f}{1}{F}{-m-2}{1}{m+1} \threej{F}{1}{f}{-m-1}{1}{m} \left( \dyad{m}{m+2} + \dyad{m+2}{m} \right),\\
            \hat{\alpha}_{I} &=& \sum_{m=-f}^{f-2} \threej{f}{1}{F}{-m-2}{1}{m+1} \threej{F}{1}{f}{-m-1}{1}{m} \left( i\dyad{m}{m+2} - i\dyad{m+2}{m} \right),\\
            \hat{\beta} &=& \sum_{m=-f}^{f} \left[ \threej{f}{1}{F}{-m}{1}{m-1}^{2} - \threej{f}{1}{F}{-m}{-1}{m+1}^{2} \right] \dyad{m}.
        \end{eqnarray}
        \label{eq:Observables}
    \end{subequations}
\end{widetext}
Taking real and imaginary parts of the quantity allows us to solve Eqs.~\eqref{eq:RotationVsKappa}. 

It should be noted that the system has several time scales associated with different processes. Here, we exclusively focus on slowly evolving terms (evolution on the time scale of $1/\Omega_{L}$), neglecting transient effects, which typically occur on the time scale $1/\Gamma$. 

In typical experiments, room-temperature atomic vapors are optically thin media, thus one can go to the limit of $\partial_z \alpha \rightarrow \Delta\alpha/L$, where $L$ is the geometric thickness of the medium. It allows us to write the polarization rotation as
\begin{widetext}
    \begin{equation}
        \Delta \alpha = \chi L e^{-\gamma t} \left[ \expval{\hat{\alpha}_{R}}\mathcal{L}_R(\Delta)\sin (2\Omega_L t) + \expval{\hat{\alpha}_{I}}\mathcal{L}_R(\Delta)\cos(2 \Omega_L t) - \expval{\hat{\beta}} \mathcal{L}_I(\Delta) \right],
        \label{eq:SimplifiedRotationAndEllipticity}    
    \end{equation}
\end{widetext}
where $\mathcal{L}_{R(I)} (\Delta)$ is the real (imaginary) part of the Lorentz profile $\mathcal{L}(\Delta)=(\Gamma - 2i \Delta)^{-1}$. As shown in Eq.~\eqref{eq:SimplifiedRotationAndEllipticity}, the polarization rotation $\Delta \alpha$ yields information about the ground-state density matrix.  Specifically, the oscillating component of the signal is associated with the Zeeman coherences, whereas the nonoscillating component is proportional to the population imbalance. This dependence naturally implies the strategy to reconstruct specific density-matrix elements.

Until this point, the motion of atoms was neglected. Yet, in room-temperature vapors, atomic motion and hence the Doppler effect need to be taken into account. It can be achieved by substituting the detuning $\Delta$, with the Doppler-shifted detuning $\Delta-\Delta_D$, $\Delta \rightarrow \Delta - \Delta_D$, and averaging the relation over the Doppler distribution $f_D(\Delta_D) = (\Gamma_D \sqrt{\pi})^{-1} \exp (-\Delta_{D}^{2} / \Gamma_{D}^{2})$ \cite{Auzinsh2010OpticallyInteractions}, where $\Gamma_D$ is the Doppler broadening. As a result, we need to rewrite the Lorentz profiles in Eq.~\eqref{eq:SimplifiedRotationAndEllipticity} by
\begin{equation}
    \mathcal{L}_{R(I)} \rightarrow V_{R(I)} (\Delta) = (\mathcal{L}_{R(I)} * f_D)_{\Delta_D}(\Delta),
    \label{eq:DopplerSubstitution}
\end{equation}
where $(.*.)_{\Delta_D}$ stays for the convolution over $\Delta_D$. One can notice that $V(\Delta)$ is the Voigt profile by its definition.
\begin{equation}
    V_{R(I)} (\Delta) = \int \mathcal{L}_{R(I)}(\Delta-\Delta_{D}) f_D(\Delta_D) d\Delta_{D},
    \label{eq:DopplerSubstitution}
\end{equation}

Unfortunately, the information accessible through Eq.~\eqref{eq:SimplifiedRotationAndEllipticity} is limited, i.e., the signal depends only on three observables $\expval{\hat{\alpha}_R}$, $\expval{\hat{\alpha}_I}$ and $\expval{\hat{\beta}}$. Thus, to determine the remaining density-matrix elements or, to be more precise, to create informationally complete positive operator-valued measure (POVM) \cite{Deutsch2010Quantum}, we introduce a set of control pulses $Q_i$. This allows to mix the population and coherences of various magnetic sublevels and extend the set of observables
\begin{equation}
    \text{Tr}\left[ \hat{X} Q_i \rho Q_i^{\dagger} \right] = \expval{Q_i^{\dagger} \hat{X} Q_i} = \expval{\hat{X'}}
    \label{eq:Control_pulses}
\end{equation}

\section{Simulations of a qutrit}
Here, we implement our algorithm on a qutrit. In a specific case, where $f=1$ and $F=0$ (such a transition occurs, for example, in $^{87}$Rb excited at the $D_2$ line), the observables \eqref{eq:Observables} take the form
\begin{subequations}
    \begin{eqnarray}
        \expval{\hat{\alpha}_{R}} =& \dfrac{1}{3} \left( \rho_{1,-1} + \rho_{-1,1} \right),\\
        \expval{\hat{\alpha}_{I}} =& \dfrac{i}{3} \left( -\rho_{1,-1} + \rho_{-1,1} \right),\\
        \expval{\hat{\beta}} =& \dfrac{1}{3} \left( \rho_{1,1} - \rho_{-1,-1} \right).
    \end{eqnarray}
    \label{eq:Observables}
\end{subequations}

To gain access to other density matrix elements, we rotate the system with the pulses of nonoscillating magnetic fields. Assuming that the pulses rotate the state around the $z$-axis by $\varphi$ and around the $y$-axis by $\vartheta$ and the pulse lengths are negligibly short, the transformed operators are given by
\begin{equation}
    \expval{\hat{X'}} = \expval{\hat{\mathcal{D}}^{\dagger}\left(0, \vartheta, \varphi \right) \hat{X} \hat{\mathcal{D}}\left(0, \vartheta, \varphi \right)},
    \label{eq:MaxtrixRotatedGeneral}
\end{equation}
where $\hat{\mathcal{D}}$ is the quantum-mechanical rotation operator \cite{messiah1962quantum}.  Using Eq.~\eqref{eq:MaxtrixRotatedGeneral}, one can calculate the relationship of rotated operator matrix elements with the elements of the initial operator
\begin{equation}
    \hat{X}'_{mn} = \sum_{b,d} e^{i\varphi \left(d-b\right)} d^{\left( 1 \right)}_{mb} \left( \vartheta \right) d^{\left( 1 \right)}_{dn} \left( \vartheta \right) \hat{X}_{bd},
    \label{eq:DMelementsRotation}
\end{equation}
where $d^{(1)}$ is the Wigner (small) $d$-matrix of the first order \cite{Sakurai2017ModernMechanics}.  To illustrate this, we consider rotation of the system by $\vartheta=\pi/2$ and $\varphi=0$, which allows us to encode information about the $\Delta m=1$ coherences in the detectable signal
\begin{equation}
\begin{split}
    \expval{\hat{\beta}'}  =&\expval{\hat{\mathcal{D}}^{\dagger}\left(0, \dfrac{\pi}{2}, 0 \right) \hat{\beta} \hat{\mathcal{D}}\left( 0, \dfrac{\pi}{2}, 0 \right)}\\ 
=&-\dfrac{1}{3\sqrt{2}} \left( \rho_{-10}+\rho_{0-1}+\rho_{10}+\rho_{01} \right).
\end{split}
\end{equation}
Similarly, it can be shown that the rotation of the state mixes the $\Delta m=2$ coherences with the population of the $m=0$ sublevel
\begin{equation}
\begin{split}
    \expval{\hat{\alpha}'_{R}}  =&\expval{\hat{\mathcal{D}}^{\dagger}\left(0, \dfrac{\pi}{2}, 0 \right) \hat{\alpha}_R \hat{\mathcal{D}}\left( 0, \dfrac{\pi}{2}, 0 \right)}\\ 
=&-\dfrac{1}{6} \left( \rho_{11}-2\rho_{00}+\rho_{-1-1}+\rho_{1-1}+\rho_{-11} \right),
\end{split}
\end{equation}
which eventually enables the full reconstruction of the ground-state density matrix.

In order to test the developed quantum-state tomography protocol, we perform numerical simulations of a given-state density-matrix evolution. On the basis of numerical simulations, we determine the polarization rotation of $x$-polarized light, which is then contaminated with white noise and used for reconstruction of the density matrix.  The reconstruction is based on fitting the polarization-rotation signals $\Delta\alpha$ with the function
\begin{equation}
    \Delta \alpha = e^{-\gamma t} \left[ A_{\alpha} \sin (2 \Omega_{L} t) + B_{\alpha} \cos (2 \Omega_{L} t ) + C_{\alpha} \right],
\label{eq:DetectedSignal}
\end{equation}
where $A_\alpha$, $B_\alpha$, and $C_\alpha$ are the fitting parameters.  In our fitting routine, the three amplitudes are the only free parameters, i.e., the Larmor frequency $\Omega_L$ and the ground-state relaxation rate $\gamma$ are known (the Larmor frequency $\Omega_L$ may be known from the strength of the magnetic field applied during probing and the ground-state relaxation rate $\gamma$ could be determined based on other measurements, such as relaxation in the dark \cite{Graf2005RelaxationCells,Chalupczak2013Radio}).  It should be noted that constraining $\Omega_L$ and $\gamma$ allows extractions of $A$, $B$, and $C$ even in the case of a poor signal-to-noise ratio (SNR).

Combining Eqs.~\eqref{eq:SimplifiedRotationAndEllipticity}, \eqref{eq:DMelementsRotation}, and \eqref{eq:DetectedSignal} enables derivation of the explicit relations between the fitting parameters and the post-pulse ground-state density-matrix elements
\begin{subequations}
    \begin{eqnarray}
        \tilde{\rho}^{\rcirclearrow}_{1-1}=(\tilde{\rho}^{\rcirclearrow}_{-11})^*&=&
        \frac{A_{\alpha}-iB_{\alpha}}{\zeta L},\\
        \tilde{\rho}^{\rcirclearrow}_{-1-1}-\tilde{\rho}^{\rcirclearrow}_{11}&=&
        \frac{C_{\alpha}}{\zeta L}.
    \end{eqnarray}
    \label{eq:FittingParameters}
\end{subequations}

As discussed above, a single fitting provides access to three density-matrix elements (compare with a technique described, for example, in Ref.~\cite{Deutsch2010Quantum}).  Other density-matrix elements can be acquired by an appropriate choice of magnetic-field pulses [Eq.~\eqref{eq:DMelementsRotation}]. In a noiseless system, three sets of pulses allow full reconstruction of the $f=1$ state.  However, the presence of noise and imperfections of the matrix manipulation (e.g., uncertainty of the pulse-rotation angles or Larmor-precession frequency) may introduce uncertainties in the reconstruction.  This may lead to a reconstruction of the matrix that does not correspond to a physical state. To accommodate for such an uncertainty, a larger set of pulses may be used, but even then the results may be nonphysical.  To address this problem, many approaches can be implemented (see, for example, Refs.~\cite{Began2005Comprehensive,Schmied2016Quantum}). Here, we use the approach based on the minimization of the distance between the reconstructed matrices and the matrix corresponding to the closest physical state. For that, we assume that the distance between matrices is defined based on the Frobenius norm 
\begin{equation}
    \delta (\rho_r) = \sum_j\text{Tr}\left[ (\rho_j-\rho_r)^{\dagger} (\rho_j-\rho_r) \right],    
\end{equation}
where $\rho_{j}$ is the $j^\text{th}$ set of ``measured'' density-matrix elements (note that only the elements associated with the three observables are determined in a single measurement, so $\rho_j$ is not completely filled matrix). Optimization with the constraints of semipositivity and normalization is achieved by parameterizing the density matrix by the Cholesky decomposition \cite{Paris2004Quantum}.

\begin{figure}[h]
    \begin{tabular}{cc}
    \includegraphics[width=0.225\textwidth]{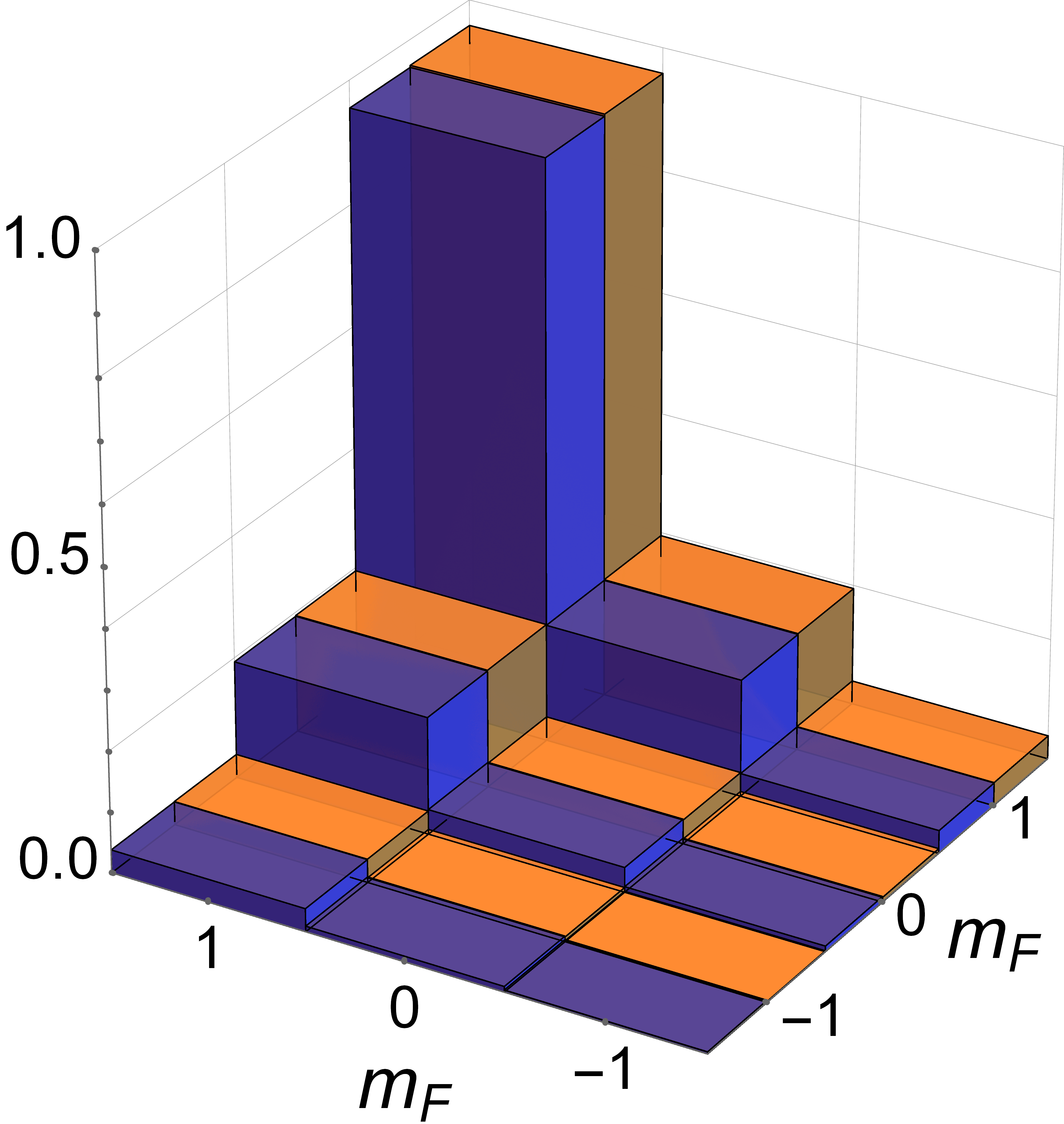} &   \includegraphics[width=0.225\textwidth]{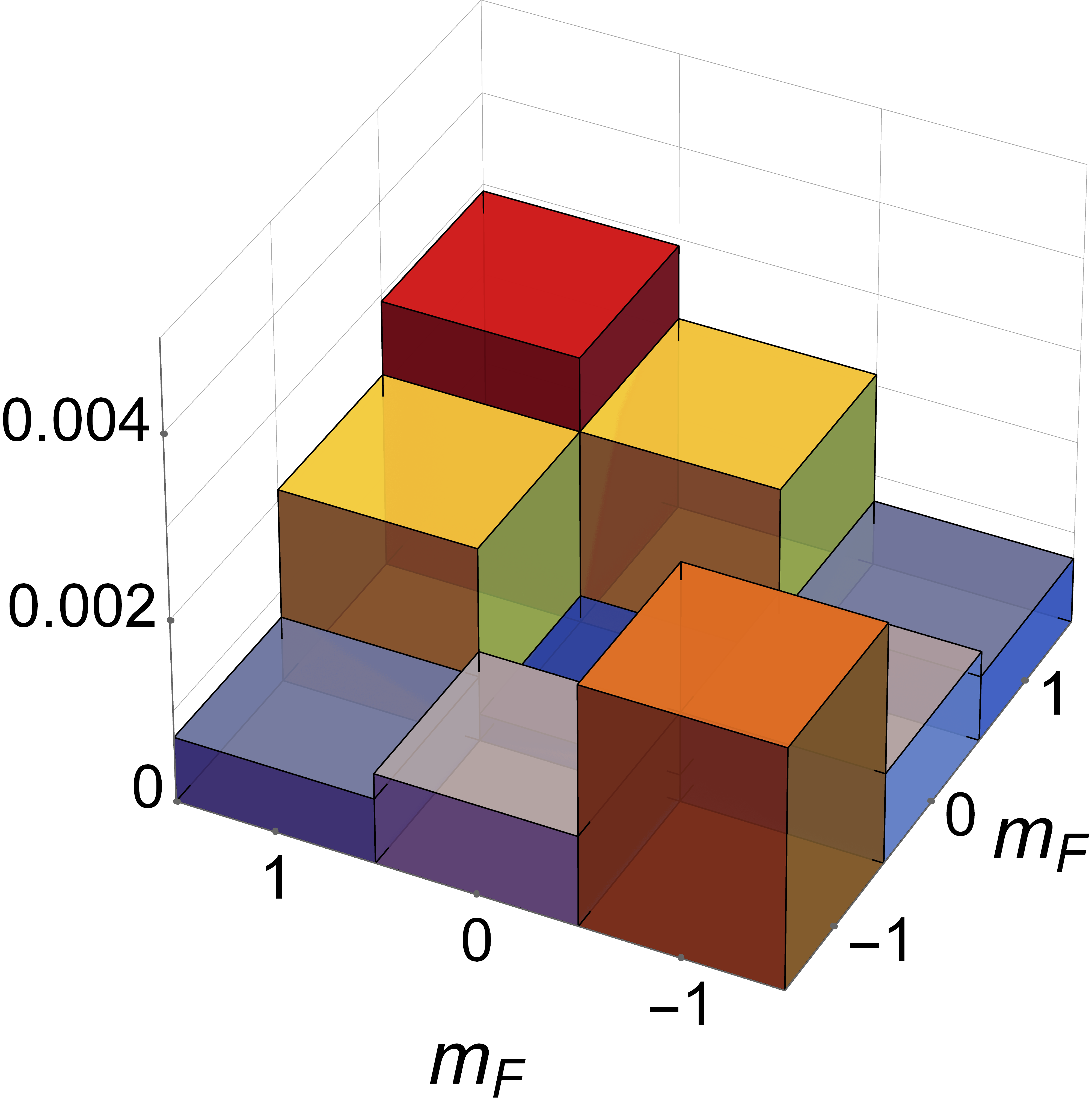} \\
    a) & b)\\[6pt]
    \includegraphics[width=0.225\textwidth]{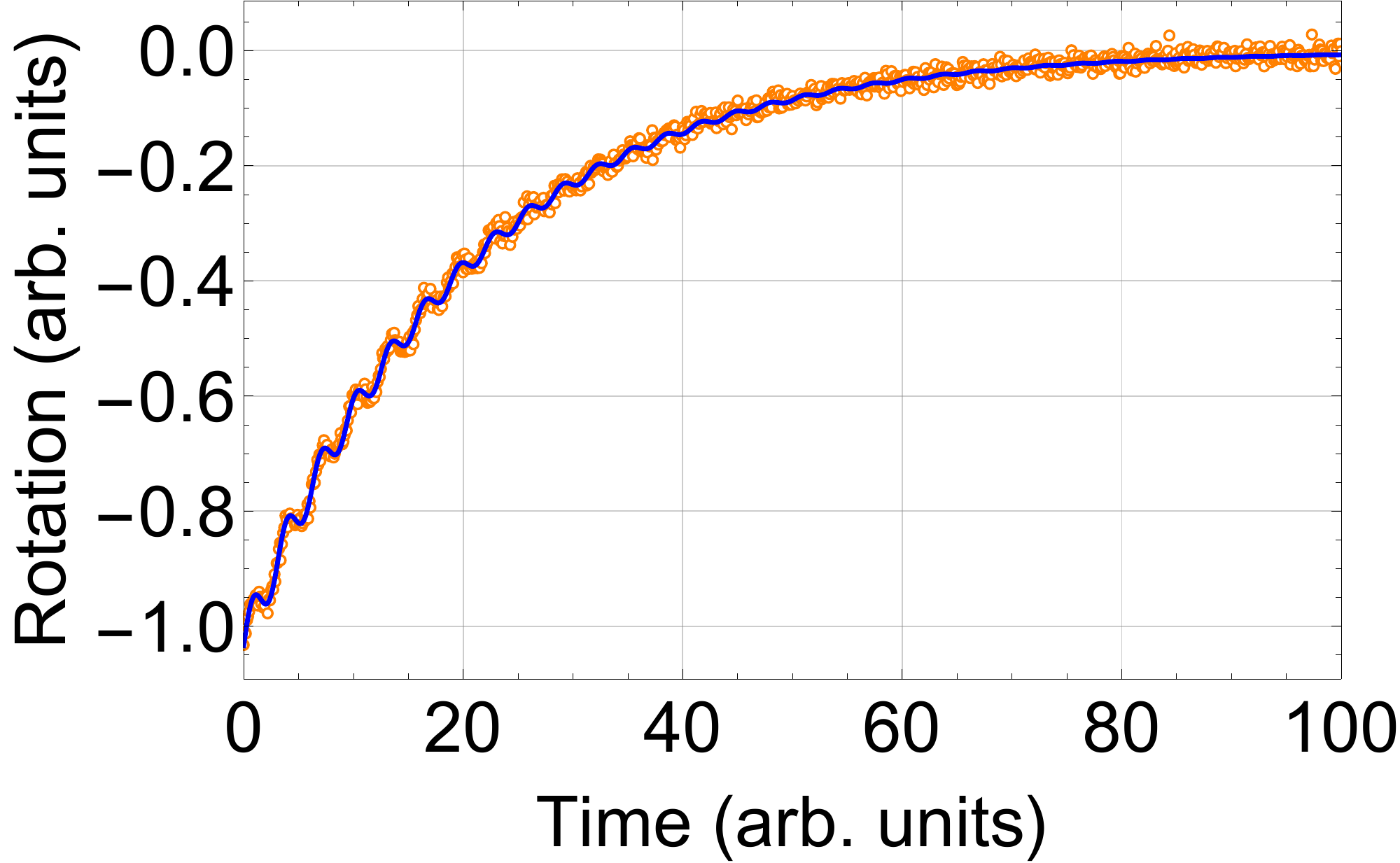} &   \includegraphics[width=0.225\textwidth]{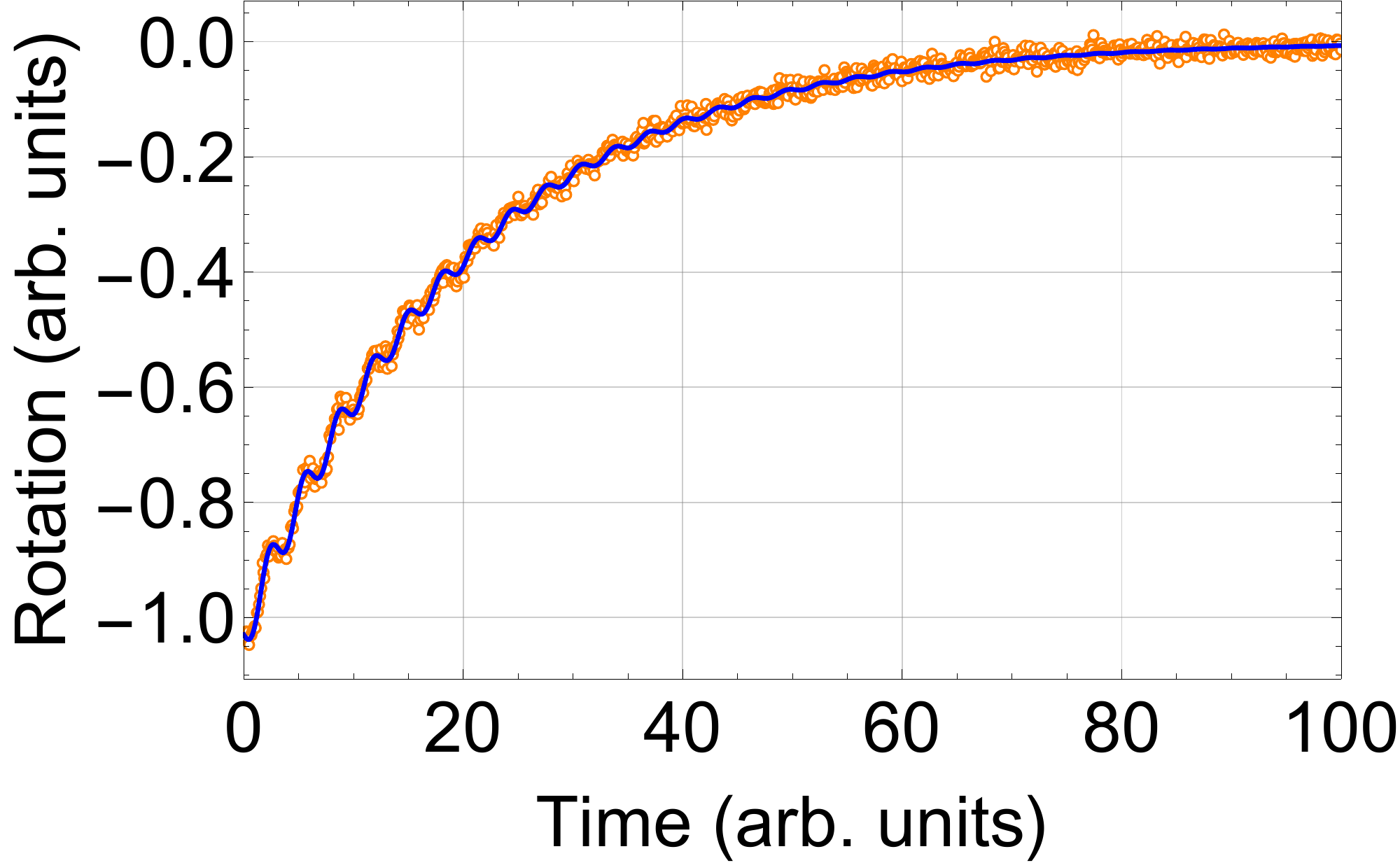} \\
    c) & d) \\[6pt]
    \includegraphics[width=0.225\textwidth]{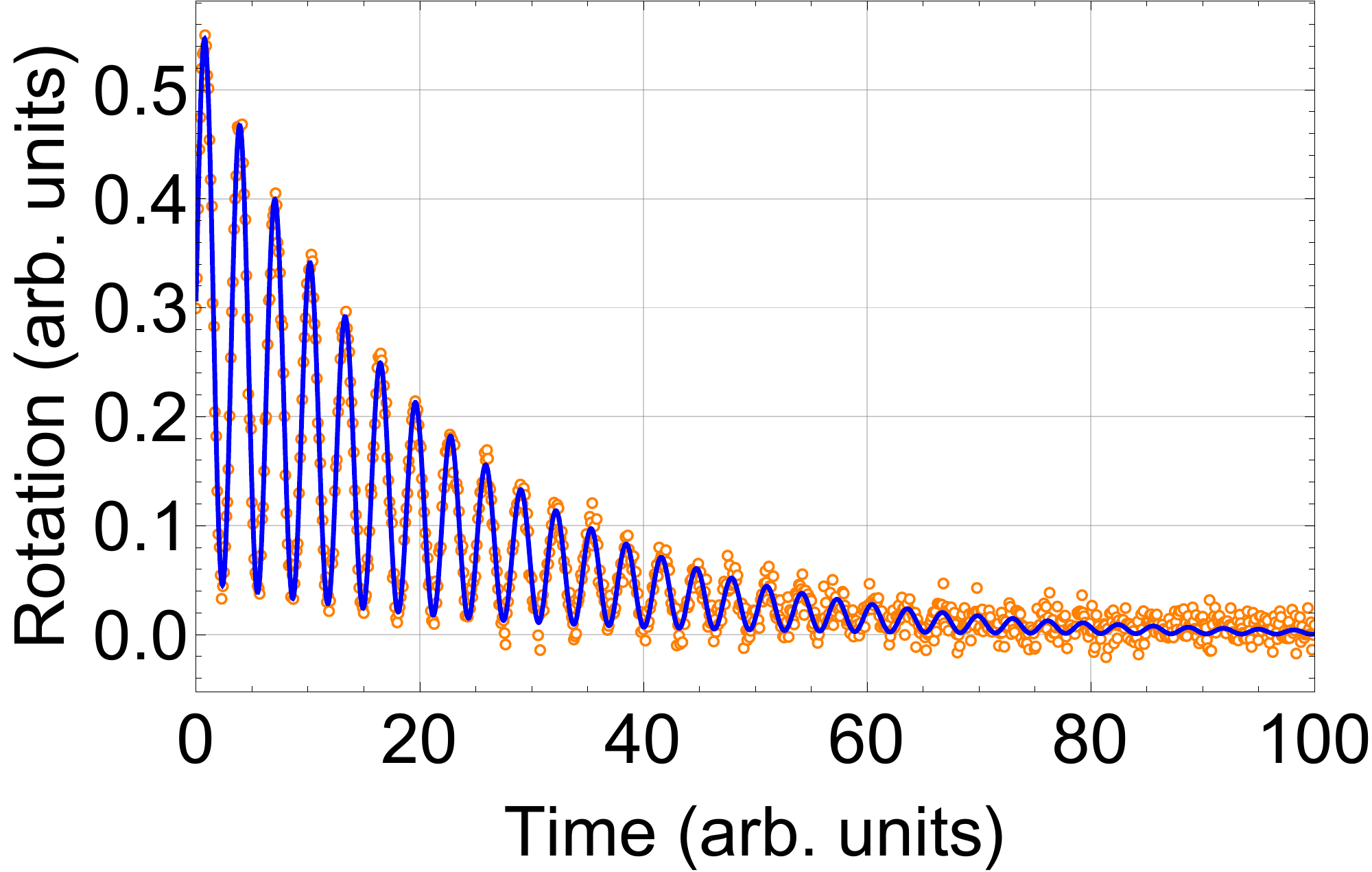} &   \includegraphics[width=0.225\textwidth]{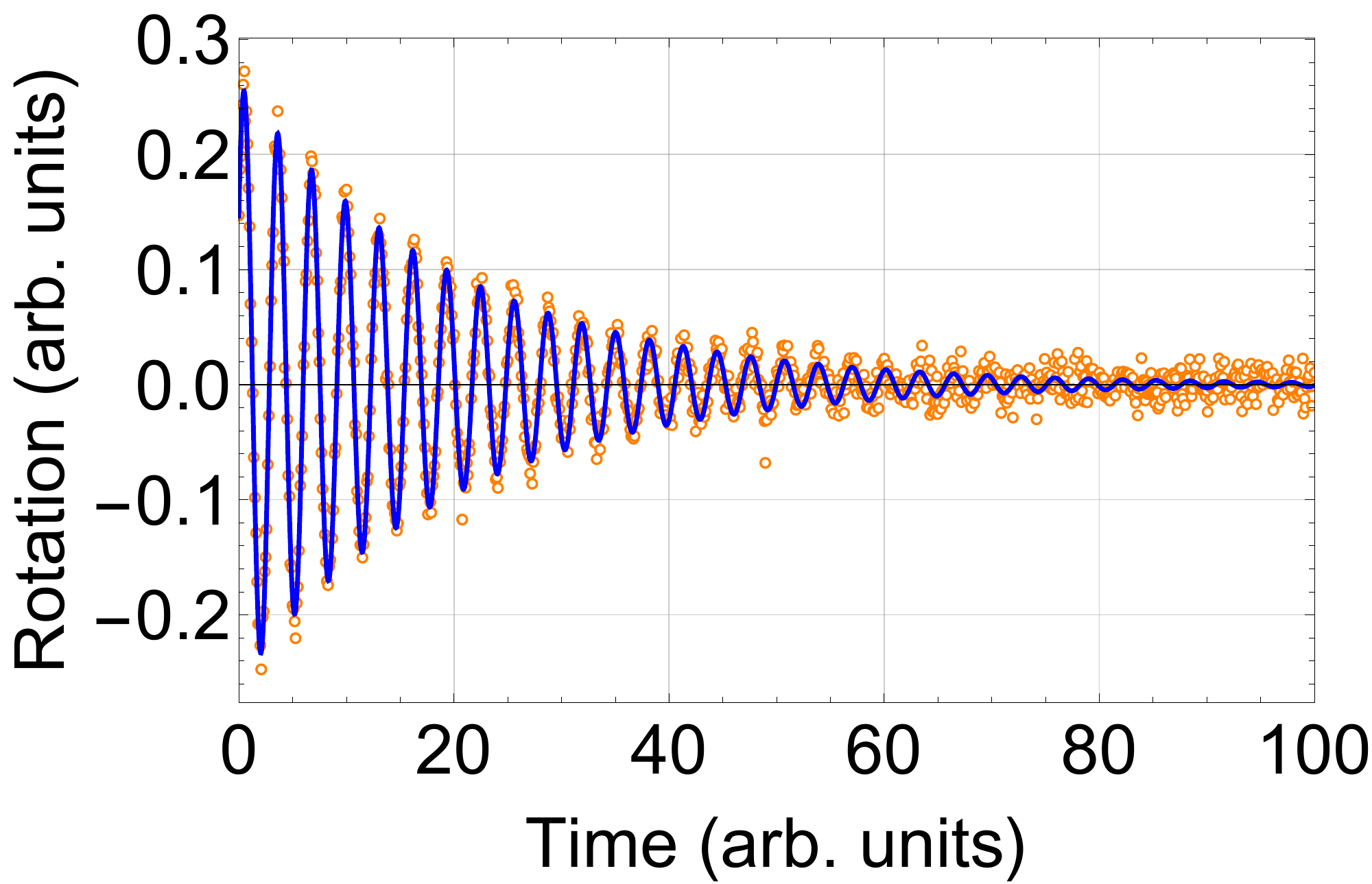} \\
    e) & f) \\[6pt]
    \end{tabular}
    \caption{(a) Magnitude of density-matrix elements of the reconstructed density matrix (blue/dark grey) of a randomly chosen pure state (orange/light grey).  (b) The difference between magnitudes of the density-matrix elements of the reconstructed and initial state. Simulated signals of polarization rotation for a set of initial magnetic-field pulses: 
    (c) $\varphi=0$ and $\theta=0$, (d) $\varphi=0$ and $\theta=\pi/2$, (e) $\varphi=\pi/2$ and $\theta=0$, and (f) $\varphi=\pi/2$ and $\theta=\pi/2$.  The simulations are performed in for the Larmor-frequency normalized set of parameters: $\Omega_R=1$,  $\Gamma=\Delta=1000$, $\gamma = 0.05$ and $\Omega_{R}=1$, $N_{at} = 10^{10}$, and $\Omega_L^{pulse}=100$, $\Omega_{L}=1$.}
    \label{fig:RandomPureState}
\end{figure}

To demonstrate the applicability of the technique, we first consider the reconstruction of a pure state.  The density matrix, corresponding to a nontrivial pure state, is shown in Fig.~\ref{fig:RandomPureState}a) (blue bars).
Polarization-rotation signals [Figs.~\ref{fig:RandomPureState}c)-f)] were calculated for a set of four arbitrarily chosen magnetic-field pulses. To mimic ``experimental'' data, we contaminated the simulated signals with white noise. To introduce a measure of the noise, we define the SNR as the ratio between the amplitude of a signal measured for a fully aligned state (see SI) to the root-mean-square of noise. In the considered case, the generated signals were burdened with noise corresponding to $\textrm{SNR}=25$. The data was then fitted using Eq.~\eqref{eq:DetectedSignal} and the density matrix was reconstructed [Fig.~\ref{fig:RandomPureState}a), orange bars].  As shown in Fig.~\ref{fig:RandomPureState}b), presenting the difference between the assumed and reconstructed density matrix, the reconstruction is reliable and the error is only a small fraction of the reconstructed values. To quantify the performance of the protocol, we calculated the fidelity $\mathcal{F}$ between the reconstructed density matrix $\rho_r$ and the matrix $\rho'$ used for the calculations \cite{Liang_2019}
\begin{equation}
   \mathcal{F} (\rho_{r},\rho') = \left( \text{Tr}\left[ \sqrt{ \sqrt{\rho_{r}} \rho' \sqrt{\rho_{r}} } \right] \right)^{2},
\end{equation}
which, in the considered case, was 0.997.

\begin{figure}[h]
    \begin{tabular}{cc}
        \includegraphics[width=0.225\textwidth]{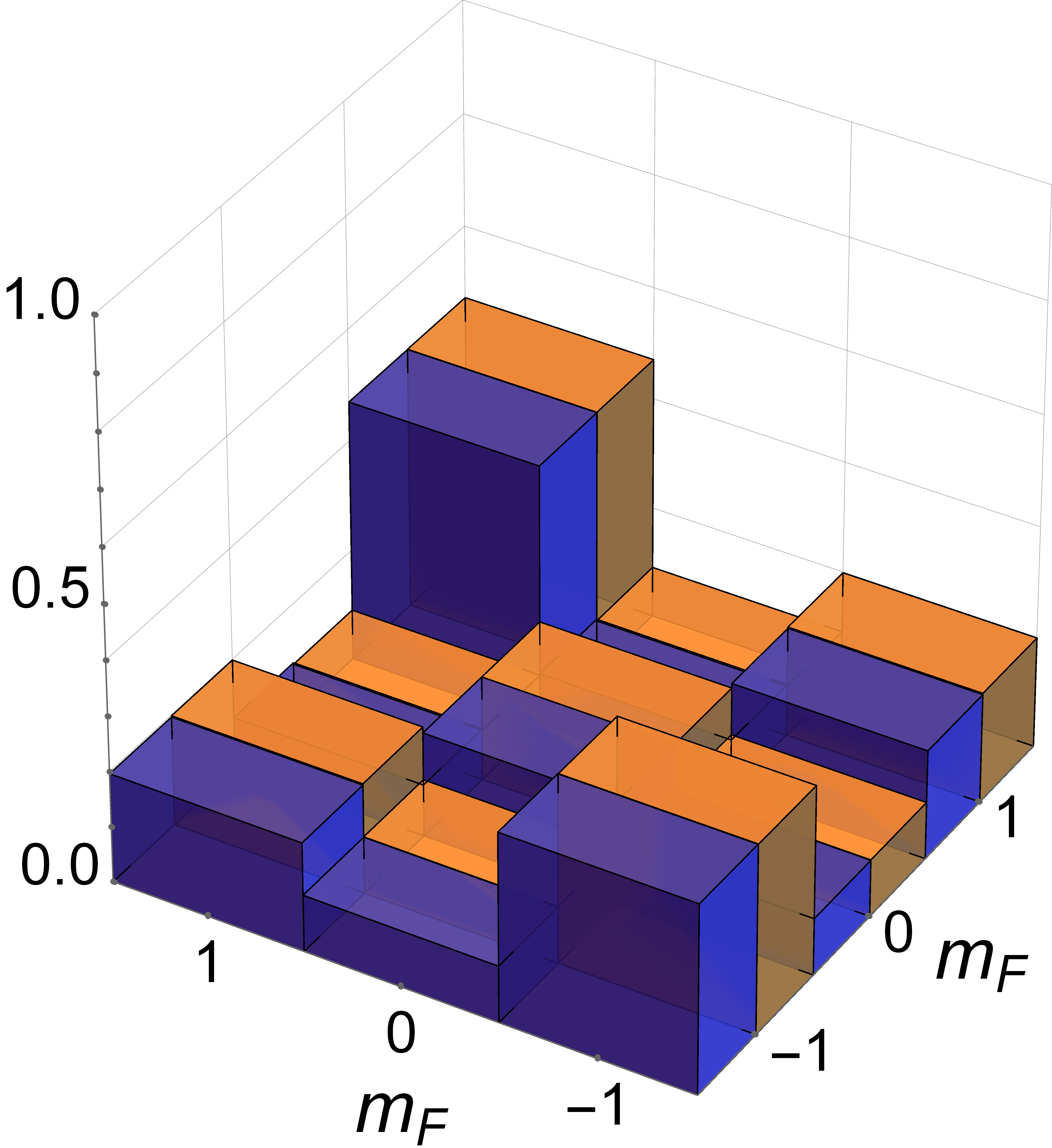} &   \includegraphics[width=0.225\textwidth]{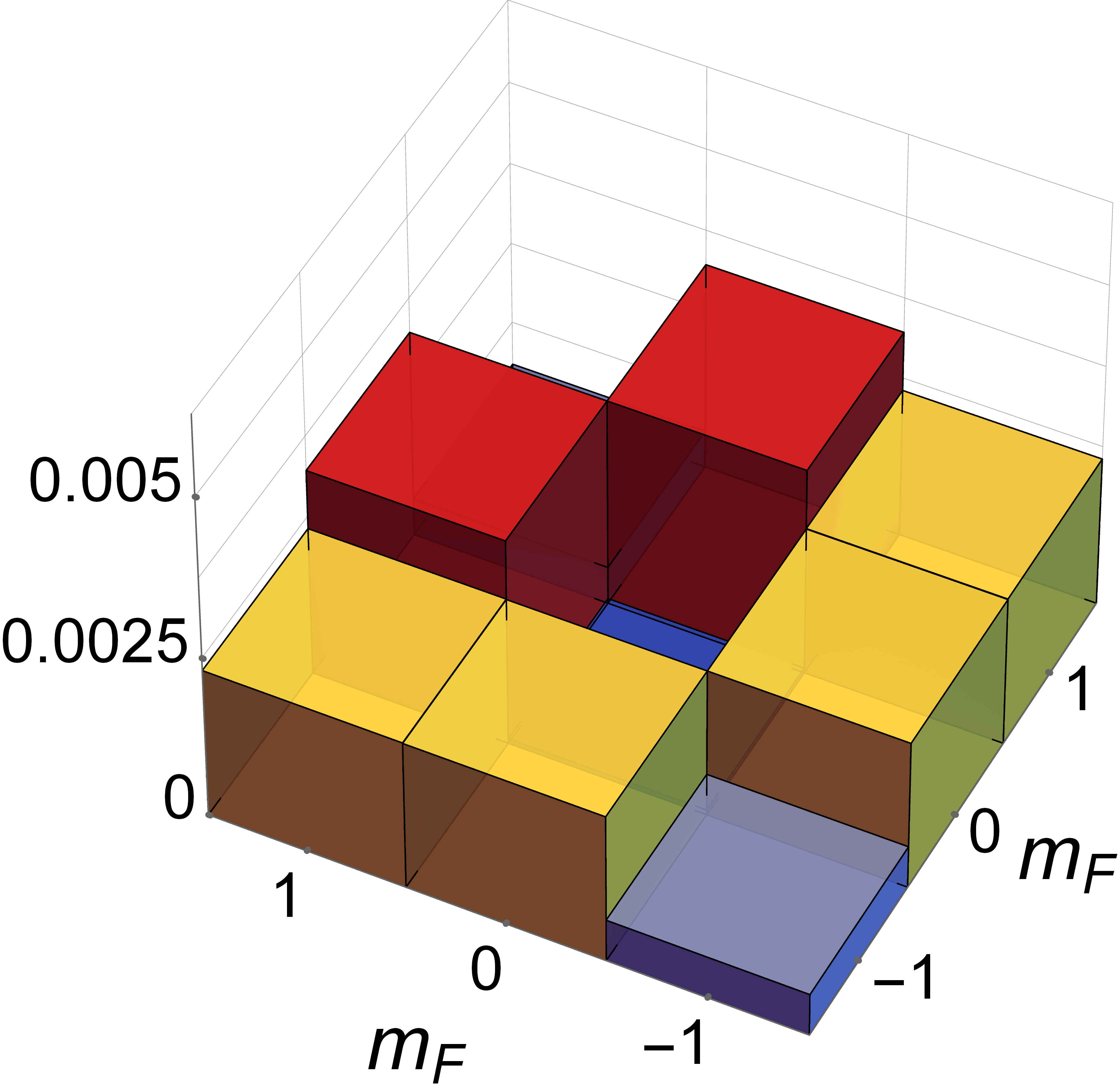} \\
        a) & b)\\[6pt]
        \includegraphics[width=0.225\textwidth]{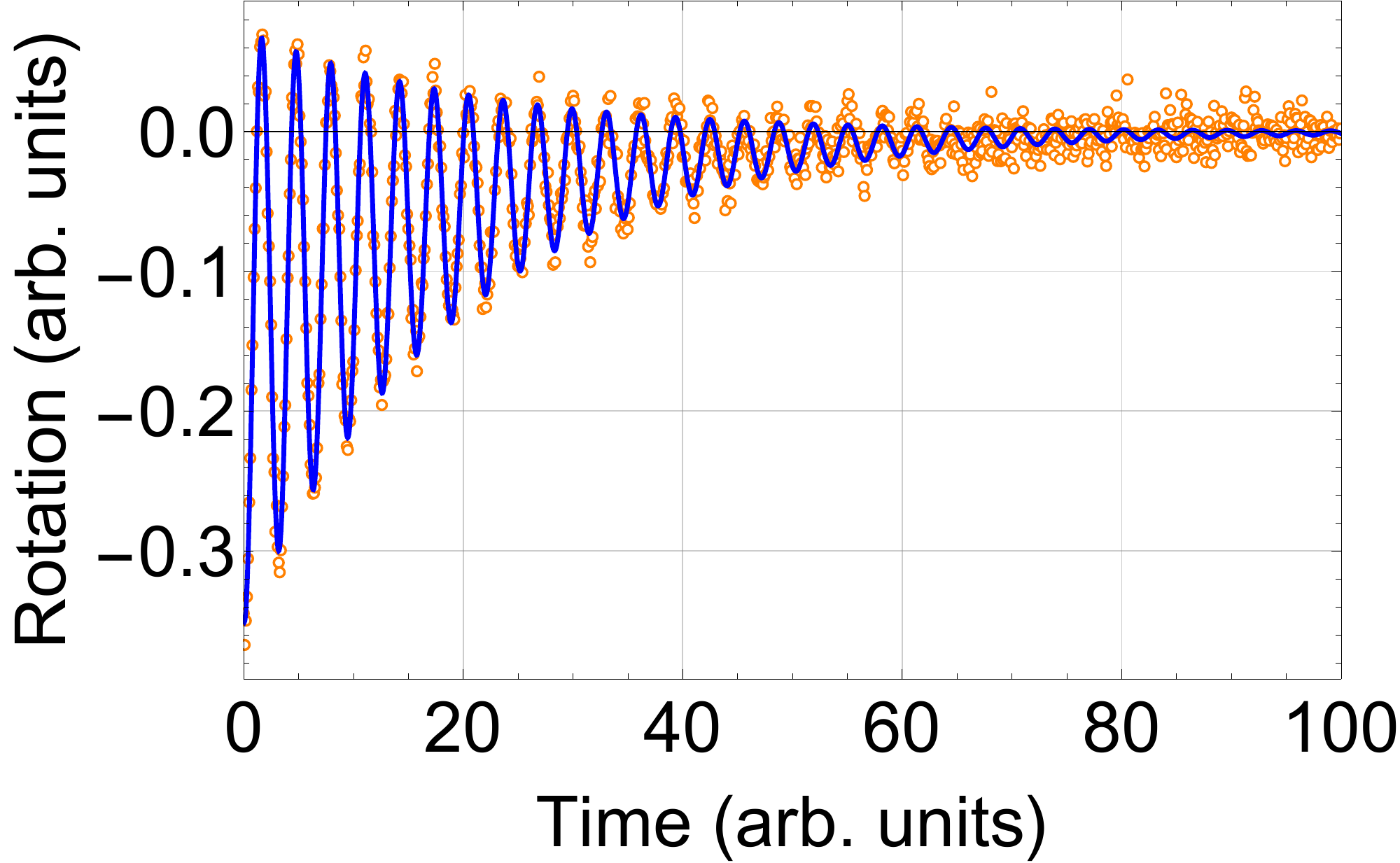} &   \includegraphics[width=0.225\textwidth]{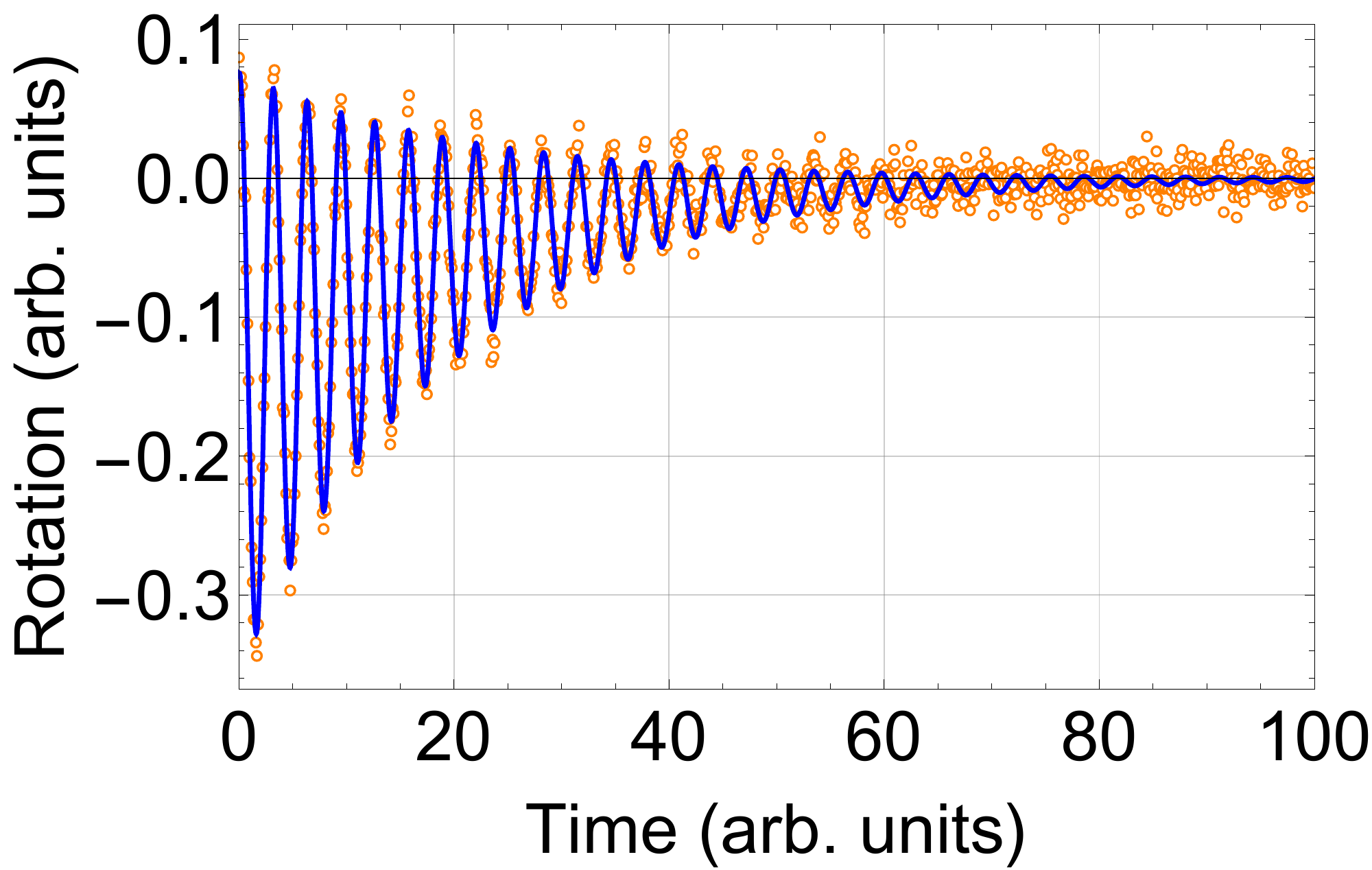} \\
        c) & d) \\[6pt]
        \includegraphics[width=0.225\textwidth]{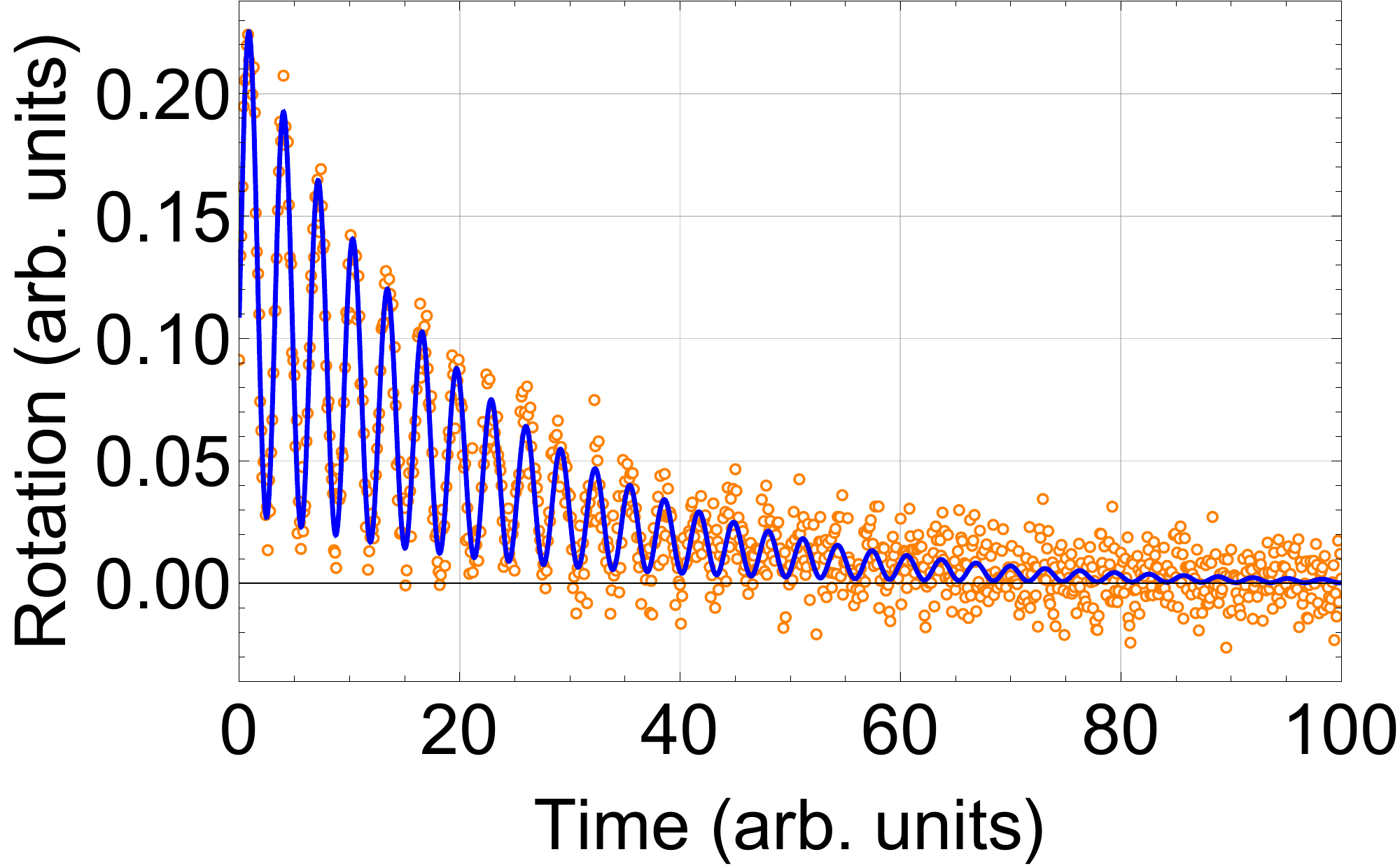} &   \includegraphics[width=0.225\textwidth]{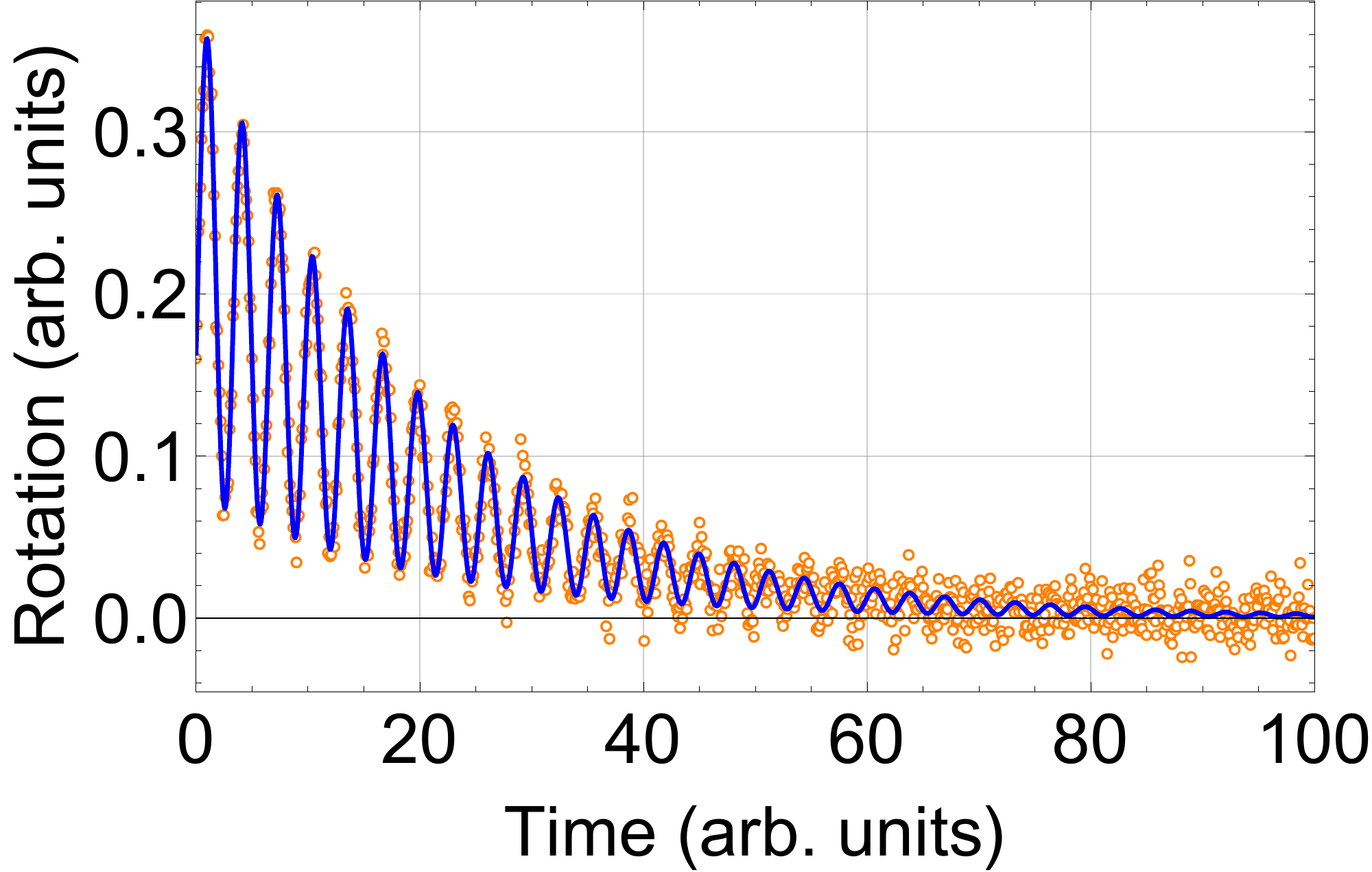} \\
        e) & f) \\[6pt]
    \end{tabular}
    \caption{Reconstruction of the partially mixed state: initial - orange/light grey and reconstructed - blue/dark grey (a) state and absolute values of differences between matrices elements (b). The rotation signals, along with the fitting, for $\varphi=0$ and $\theta=0$ (c), $\varphi=0$ and $\theta=\pi/4$ (d), $\varphi=\pi/4$ and $\theta=0$ (e), and $\varphi=\pi/4$ and $\theta=\pi/4$ (f). The simulation parameters identical as in Fig.~\ref{fig:RandomPureState}.}
    \label{fig:RandomMixedState}
\end{figure}

Next, we analyzed the reconstruction of a partially mixed state (Fig.~\ref{fig:RandomMixedState}).
Similarly as before, the rotation signals were simulated and white noise was added to the data.  A noticeable difference between this and the previous result is the amplitude of the observed signals, being smaller for the mixed state.  In fact, one can formulate the general observation that the more mixed the state, the lower the amplitude of the signal. The fitting of the reconstruction of the state was 0.996.

An important question is the role of SNR and uncertainty of the rotation pulses in the fidelity of the state reconstruction.  To address this question, we analyze the fidelity of the reconstruction of pure, partially mixed (an averaged purity of~0.6), and fully mixed (unpolarized) states versus SNR and uncertainty of rotation induced by DC pulses. For each SNR or angle uncertainty, the reconstruction was performed 100 times with a set of four control pulses (four pairs of rotation angles), rotating the matrix around the $z$- and $y$-axes at random angles. Randomization of rotation angles was introduced to avoid trapping of the reconstruction in local minima. The fidelity of the reconstruction for the pure and partially-mixed states was averaged over ten different states, which were obtained from the stretched state by a unitary transformation, generated from the Haar measure~\cite{Zyczkowski2011GeneratingMatrices}.

\begin{figure}[h]
    \begin{tabular}{cc}
        \includegraphics[width=0.225\textwidth]{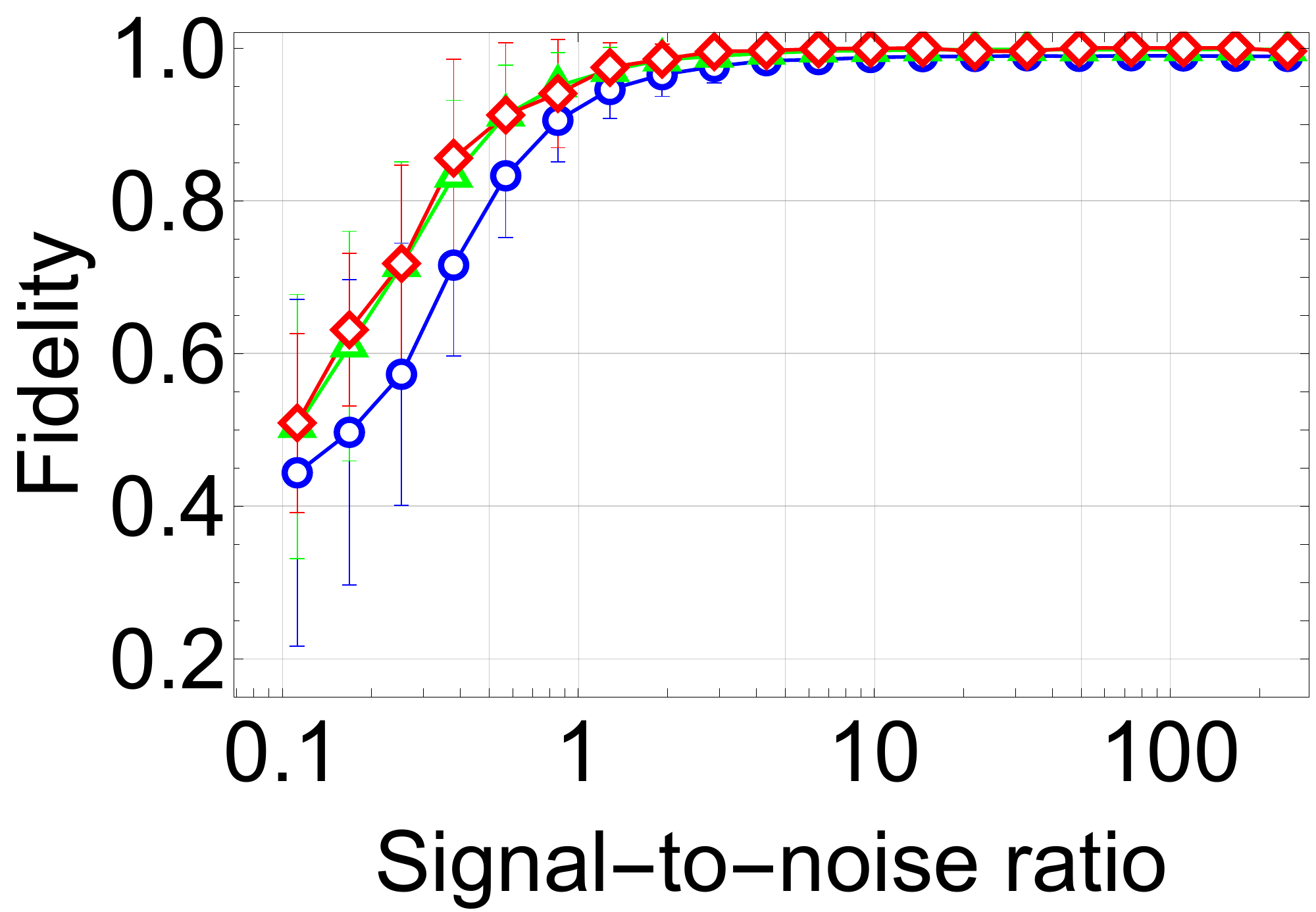} &   \includegraphics[width=0.225\textwidth]{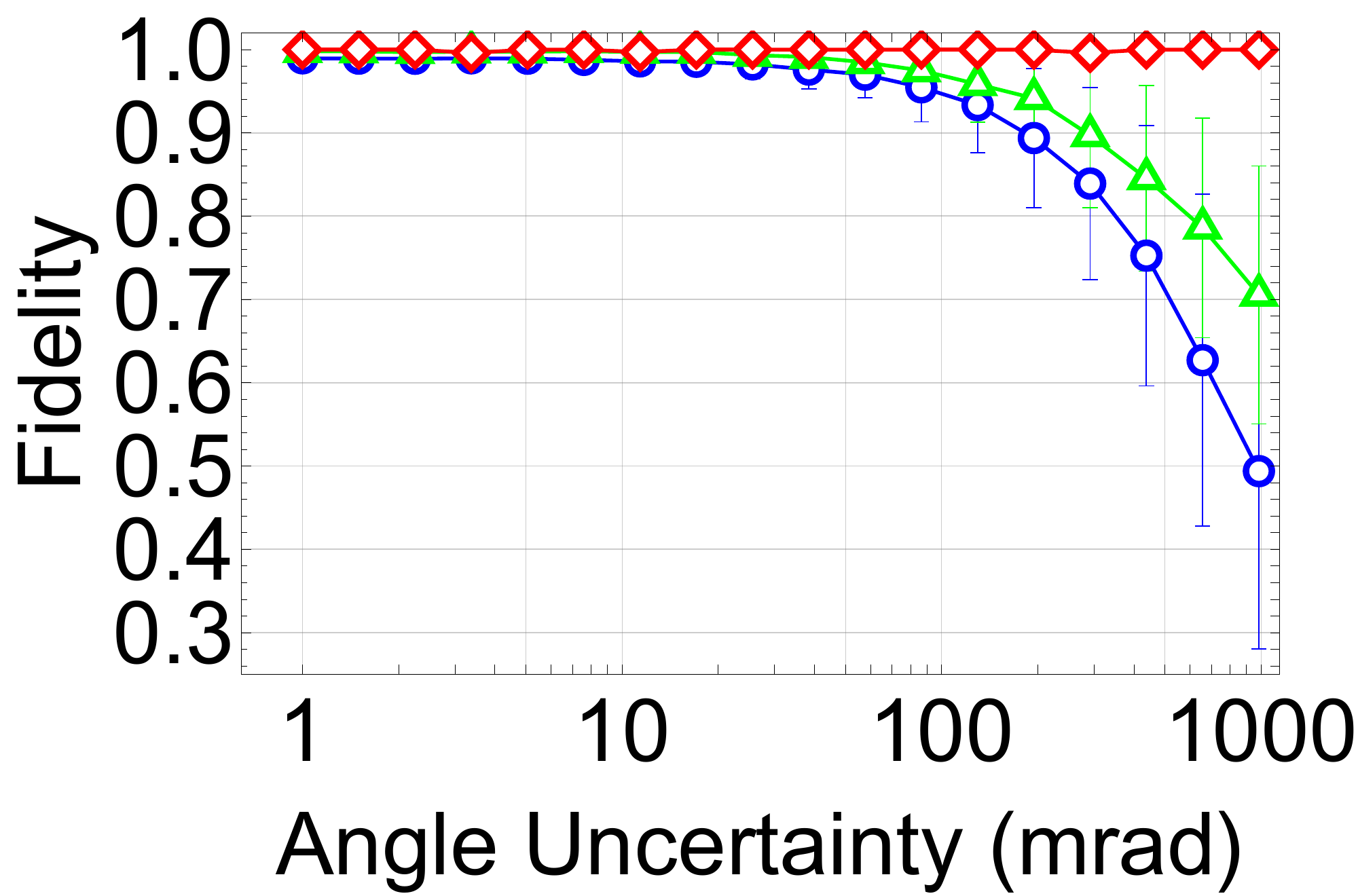} \\
        a) & b)
    \end{tabular}
    \caption{Fidelity of the state reconstruction versus SNR (a) and uncertainty of the magnetic-field pulse rotation (b).  The results are shown for a pure states (blue circles), partially mixed - purity $\approx 0.6$ (green triangles) and fully mixed state (red diamonds).}
    \label{fig:FidelityVsSNR}
\end{figure}

 To estimate the average value and the variance of the fidelity for each set of parameters, we simulate a 1000 reconstruction (10 states $\times$ 100 different sets of random rotation angles and compare the reconstructed state with the assumed one by calculating the fidelity. We plot the achieved fidelities for a set of simulation parameters on a histogram and fit the beta distribution \cite{Enriquez2018EntanglementStates}. The choice of beta distribution was dictated by the fact that fidelity is limited to the range between zero and unity, so modeling it by the normal distribution would not yield reasonable conclusions. After fitting, we calculate the expectation value and the variance of this fitted distribution and treat them as an average value and variance of fidelity for the chosen set of parameters. The results, shown in Fig.~\ref{fig:FidelityVsSNR}, reveal that already with a single set of four measurements, each with SNR at a level of 1, the fidelity exceeds 0.9 and at a level of 10, the fidelity is close to unity.
As described above, such high reconstruction fidelity results from constraining the Larmor frequency $\Omega_L$ and relaxation/decay rate $\gamma$ of the signal and, therefore, improving the fitting precision of $A_\alpha$, $B_\alpha$, and $C_\alpha$.  

Another question concerns the role of rotation uncertainty.  We showed with the simulations that rotation uncertainty up to several tens of milliradians has almost no effect on the fidelity of the reconstruction. For larger uncertainties, the fidelity starts to deteriorate (with the distinct exception of the thermal state).  Nevertheless, the precision of the rotation at the level of a few degrees, which is easily achieved experimentally, ensures a good reconstruction of the state.

\begin{figure}[h]
    \begin{tabular}{cc}
         \includegraphics[width=0.225\textwidth]{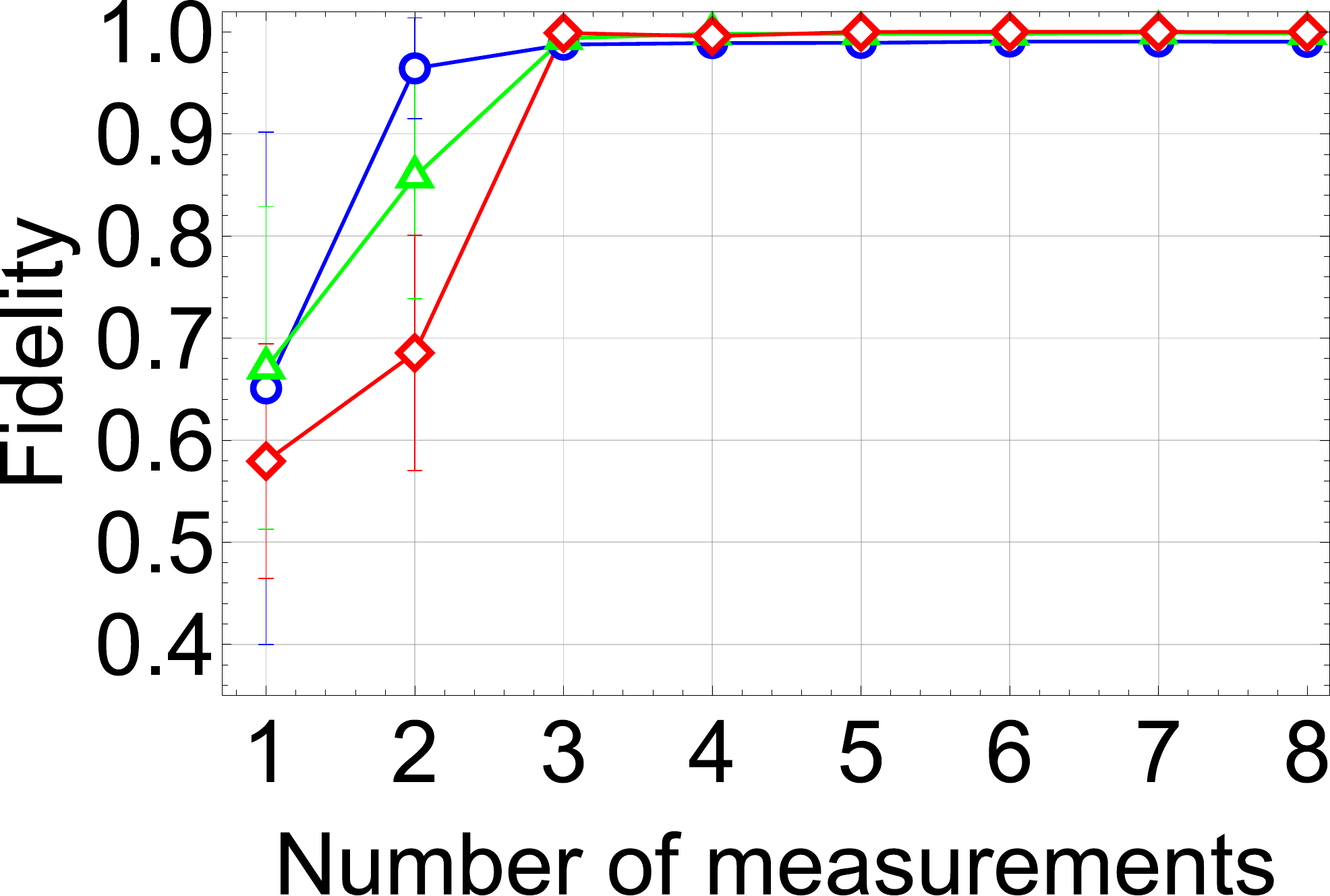} & \includegraphics[width=0.225\textwidth]{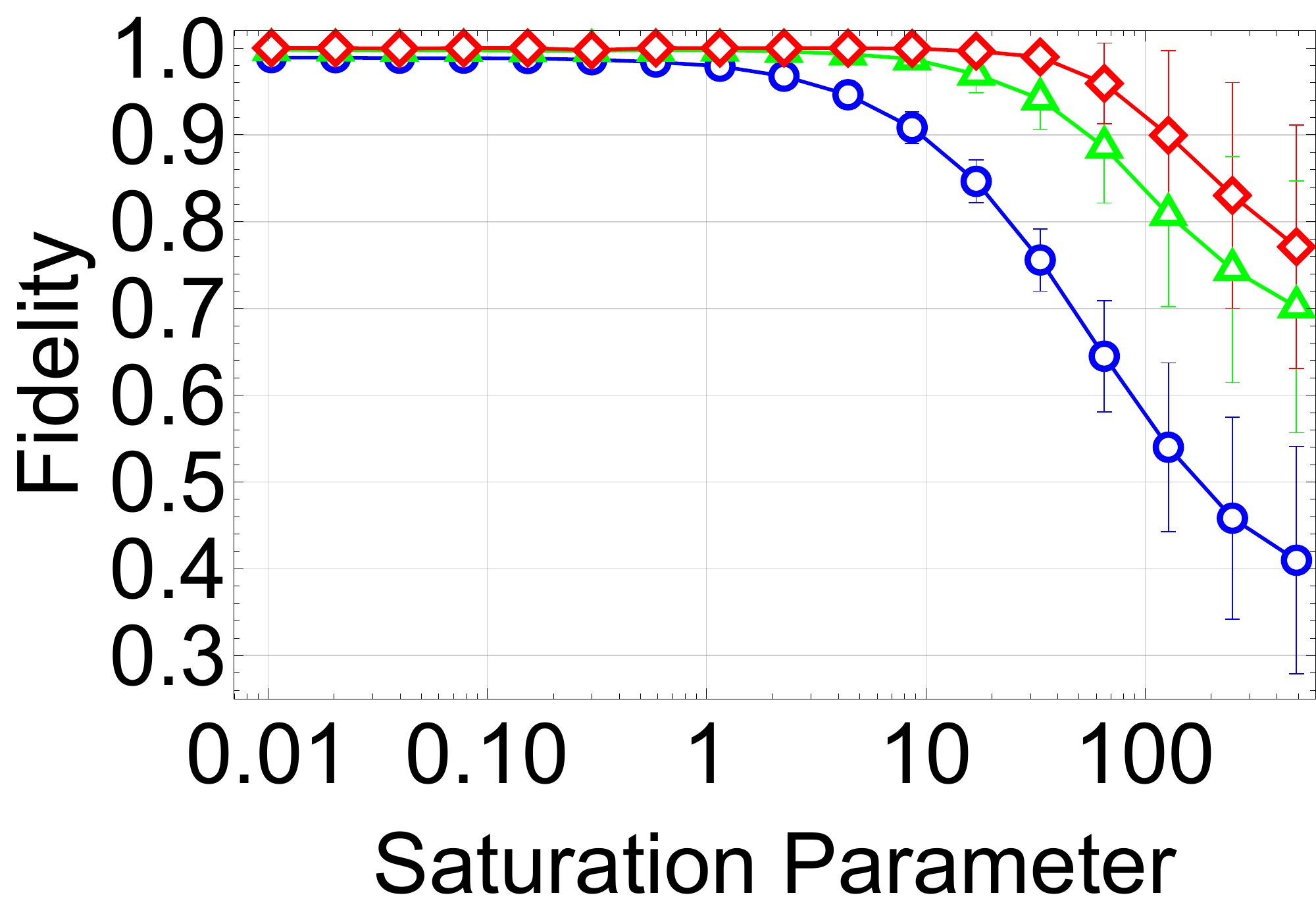}\\
        a) & b)
    \end{tabular}
    \caption{Fidelity of the reconstruction versus (a) number of measurements and (b) strength of the light-matter coupling, expressed as the saturation constant $\kappa_{2}$, for pure (blue circles), mixed (green triangles), and thermal (red diamonds) states. The calculations were performed for the parameters same as in Fig.~\ref{fig:FidelityVsSNR} with SNR $\approx 25$ and no uncertainty in the rotation.}
    \label{fig:FidelityVsNumber}
\end{figure}

From the perspective of practical implementation of the protocol \cite{Kopciuch2022}, an important question concerns the number of measurements needed for a reliable reconstruction of a quantum state. Generally, in the considered system, one needs to measure at least three polarization-rotation signals to fully reconstruct the density matrix.  However, as shown in Fig.~\ref{fig:FidelityVsNumber}a), for even lower number of measurements, partial reconstruction is possible, although its fidelity is small.
In principle, the higher the number of measurements, the more precise the reconstruction.  This becomes particularly important in noisy media, where low signal quality needs to be overcome by more measurements, so that minimization, being the last stage of the algorithm, can be performed more precisely. 

A different contribution to the uncertainty of the reconstruction comes from the action of the probing beam.  For our considerations, we have assumed that the action of probe light is so small that it is negligible.  Generally, this is not true, and the probe action can manifest itself at various levels.  For instance, even in a simple case of classical light, there is nonzero probability of the optical excitation by an off-resonant (probe) beam.  This manifests itself as optical pumping, which modifies the reconstructed state and hence decreases the fidelity of the reconstruction. This effect is shown in Fig.~\ref{fig:FidelityVsNumber}b), where fidelity is measured versus the probe-light saturation parameter $\kappa_{2} = \Omega_{R}^{2}/(\Gamma \gamma)$ \cite{Auzinsh2010OpticallyInteractions}.

For small $\kappa_{2}$, the pumping is inefficient, so the probe light does not affect the state. However, the higher the probe-light intensity, the more probable the excitation is. In turn, this pumping with the probe leads to deterioration of the fidelity of the reconstruction.  This effect is reduced in paraffin-coated cells, where the atoms, moving freely through the cell, enter and exit the light beam, which effectively averages the perturbation across the entire cell \cite{Zhivun2016VectorCells}. It allows us to achieve an effective saturation parameter roughly two orders of magnitude smaller than that in a cold atomic ensemble \cite{Zhivun2016VectorCells}.

\section{Conclusions}
In this work, we described the technique for reconstructing a collective density matrix of an arbitrary quantum state. By deriving explicit relationships between polarization rotation and ellipticity change with the ground-state density-matrix elements, we demonstrate the ability of the full reconstruction of an unknown state. This technique was then analyzed in the context of the reconstruction of the quantum state of the long-lived $f=1$ state that can act as a qutrit. With the analysis, the robustness of the technique against noise, experimental uncertainties, and the number of measurements was demonstrated. 

In the near future, we plan to use the technique to reconstruct a collective density matrix of the $f=1$ ground state of room-temperature $^{87}$Rb vapor contained in a paraffin-coated cell \cite{Kopciuch2022}.  The reconstruction of the state, initially generated and modified using a sequences of light, as well as, static and oscillating magnetic-field  pulses, will be performed via the detection of polarization rotation linearly polarized off-resonant light. The analysis will offer the first step in the practical implementation of the technique.

We plan to further expand the experimental research. In this context, however, an important question concerns the ability to experimentally reconstruct a state of an angular momentum greater than 1. As shown in this article, in principle the technique offers such a capability. However, the experimental problem is the degeneracy of the evolution frequency of $\Delta m=2$ coherences. We plan to approach it via lifting the degeneracy either by using the non-linear Zeeman effect \cite{Pustelny2011Tailoring} or the AC Stark effect \cite{Deutsch2010Quantum}. Both lead to energy shifts of magnetic sublevels, which remove the coherence degeneracy and thus differentiate contributions to the signal from different coherences. Despite the technical challenges associated with turning the fields on and off, it is believed that through such an approach, the reconstruction of an arbitrary quantum state can be performed. Specifically, we plan to implement the reconstruction of quantum states of both hyperfine levels of rubidium atoms.

\section{Acknowledgements}
The authors thank Jan Kołodyński and Adam Miranowicz for stimulating discussions. The authors acknowledge the support of the National Science Centre, Poland within the SONATA BIS programme (grant number 2019/34/E/ST2/00440).

\section{Appendix A: Derivation of relations}

\subsection{Semiclassical light-atom interaction}

In the semi-classical approximation, the relation between light and atomic properties is governed by the Maxwell equation \cite{Auzinsh2010OpticallyInteractions}:
\begin{equation}
    \nabla^{2}\mathbf{E} - \dfrac{1}{c^{2}}\partial^{2}_{t} \mathbf{E} = \mu_{0}\partial^{2}_{t} \mathbf{P},  
    \label{eq:MaxwellEquation}
\end{equation}
where $\mathbf{E}$ is the electric field of light and $\mathbf{P}$ is the polarization of the atoms.  Taking a generic form of light
\begin{widetext}
    \begin{equation}
    \mathbf{E} = E_{0}\text{Re}\Bigg\{ e^{i( kz -\omega t + \phi ) } \Big( \cos(\epsilon) \big[ \cos(\alpha) \mathbf{x} + \sin(\alpha) \mathbf{y} \big] + i\sin(\epsilon) \big[ \cos(\alpha) \mathbf{y} - \sin(\alpha) \mathbf{x} \big] \Big)  \Bigg\},
    \label{eq:ElectricFieldOfLight}
    \end{equation}
\end{widetext}
where $E_{0}$, $\phi$, $\epsilon$, $\alpha$, $\omega$ and $k$ are the amplitude, phase, ellipticity, polarization-orientation angle, frequency, and wavenumber of light, respectively, and assuming that the polarization can be written as
\begin{equation}
    \mathbf{P} = \text{Re}\Bigg\{ e^{i (kz - \omega t +\phi)} \Big[ (P_{(1)} - i P_{(2)}) \mathbf{x} + (P_{(3)} - i P_{(4)}) \mathbf{y}\Big] \Bigg\}
    \label{eq:PolarizationMedium}
\end{equation}
where $P_{(i)}$ are the respective quadrature components of the atomic polarization, explicit relations can be drawn between the specific properties of light and the given polarization quadratures. For convenience, we introduce a complex atomic polarization vector $\mathbf{\mathcal{P}}$, the components of which are defined by Eq.~\eqref{eq:PolarizationMedium}. Substituting Eqs.~\eqref{eq:ElectricFieldOfLight} and \eqref{eq:PolarizationMedium} into Eq.~\eqref{eq:MaxwellEquation} and neglecting the second-order derivatives enable one to derive relations for polarization rotation and degree of ellipticity on the quadratures $P_{(i)}$.  Taking linearly polarized light ($\epsilon=0$) along the $y$-axis ($\alpha = \pi/2$), the polarization rotation, ellipticity, phase change and absorption, defined as partial derivatives over distance, are given by
\begin{subequations}
    \begin{eqnarray}
        \partial_{z} \alpha =& -\dfrac{\omega}{2 \varepsilon_{0} E_{0} c} P_{(2)},\\
        \partial_{z} \epsilon =&  -\dfrac{\omega}{2 \varepsilon_{0} E_{0} c} P_{(1)},\\
        \partial_{z} \varphi =& \dfrac{\omega}{2 \varepsilon_{0} E_{0} c} P_{(3)},\\
        \dfrac{1}{E_0}\partial_{z} E_0 =& \dfrac{\omega}{2 \varepsilon_{0} E_{0} c} P_{(4)}. 
    \end{eqnarray}
    \label{eq:RotationEllipticityPolarization}
\end{subequations}

Using the atomic-polarization components, Eqs.~\eqref{eq:RotationEllipticityPolarization} can be written as
\begin{subequations}
    \begin{eqnarray}
        \mathcal{P}_x &=& P_{(1)}+iP_{(2)} =  -\dfrac{2\varepsilon_0 E_0 c}{\omega}\left( \partial_z \epsilon - i\partial_z \alpha \right)\\
        \mathcal{P}_y &=& P_{(3)}+iP_{(4)} = \dfrac{2\varepsilon_0 E_0 c}{\omega}\left( \partial_z \varphi - i\dfrac{1}{E_0}\partial_z E_0 \right)
    \end{eqnarray}
    \label{eq:RotationEllipticityPolarization1}
\end{subequations}
The atomic polarization can be derived as an expectation value of the dipole operator $\hat{\mathbf{d}}$
\begin{equation}
    \mathbf{P} = N_{at}\Big< \hat{\mathbf{d}} \Big> =N_{at} \text{Tr}\Big[ \rho \hat{\mathbf{d}} \Big],
    \label{eq:PolarizationFM}
\end{equation}
where $\rho$ is the density matrix, $N_{at}$ is the atomic number density and, $\Big< \Big>$ denotes the expectation value. It is more convenient to calculate the components of the atomic polarization in the rotation basis. To do so, the dipole operator is transformed to the rotating basis
\begin{equation}
    \hat{d}^q = U^{*}_{qi}\hat{d}_i,    
\end{equation}
where $U$ is the transformation to the rotation basis \cite{Auzinsh2010OpticallyInteractions} and then the atomic-polarization components take the form
\begin{equation}
\begin{split}
    P_{(i)} &= 2N_{at} \text{Re} \left\lbrace \sum_q \sum_{m\mu}  \rho_{\mu m} U_{iq}^{\mathrm{T}} d^{q}_{m\mu} \right\rbrace\\
    &=\text{Re}\left \lbrace \sum_{q} U_{iq}^{\mathrm{T}} \left(2 N_{at}\sum_{m\mu} d^{q}_{m\mu} \rho_{\mu m} \right) \right \rbrace\\
    &=\text{Re}\left \lbrace \mathrm{e}^{-i \omega t} \sum_{q} U_{iq}^{\mathrm{T}} \left(2 N_{at}\sum_{m\mu} d^{q}_{m\mu} \tilde{\rho}_{\mu m} \right) \right \rbrace\\
    &=\text{Re}\left \lbrace \mathrm{e}^{-i \omega t} \sum_{q} U_{iq}^{\mathrm{T}} \mathcal{P}^{q} \right \rbrace\,.\\
\end{split}
\end{equation}
In the first equity factor 2 and the real part comes from the summation over complex conjugated terms in trace in \eqref{eq:PolarizationFM}. In the third equity, we transform the density matrix to the rotating frame:
\begin{equation}
\label{eq:RWA}
    \rho = \hat{M} \tilde{\rho} \hat{M}^{\dagger} = 
    \left( \begin{array}{ll}
    (\tilde{\rho}_{mn}) & (\tilde{\rho}_{m\mu}) \mathrm{e}^{i\omega t}  \\
    (\tilde{\rho}_{\mu m}) \mathrm{e}^{-i \omega t} & (\tilde{\rho}_{\mu\nu})  \\
    \end{array} \right),
\end{equation}
where $\hat{M}=\hat{P}_g + \mathrm{e}^{i \omega t} \hat{P}_e$ is the transformation into the rotating frame, with $P_{g(e)}$ beginning the projection operator into the ground- and excited-state subspace. $(\tilde{\rho}_{mn})$ and $(\tilde{\rho}_{\mu \nu})$ are the submatrices corresponding to the ground and excited states, and $(\tilde{\rho}_{m \mu})$ denotes the submatrix describing the optical coherences between the states. To keep text more clear, we hold the convention that for the magnetic quantum number the Latin letter denotes the ground state, while the Greek stands for the excited state. 

Transferring the system to the rotating frame allows us to obtain a time-dependent version of Eq.~\eqref{eq:PolarizationMedium}. Using the Wigner-Eckart formula \cite{messiah1962quantum} and recalculating the reduced matrix element of the dipole operator from the total angular momentum to the total electronic angular momentum, one can show that
\begin{widetext}
    \begin{equation}
        \mathcal{P}^{q} = 2N_{at} (-1)^{j+f+I+F+1} \sqrt{(2f+1)(2F+1)} \expval{j \norm{d^{(J)}} J} \sixj{j}{f}{I}{F}{J}{1} \left[ \sum_{m\mu} (-1)^m \threej{f}{1}{F}{-m}{q}{\mu} \tilde{\rho}_{\mu m} \right]
        \label{eq:complex_polarization}
    \end{equation}
\end{widetext}

This allows us to calculate an explicit formula for polarization rotation and ellipticity change.
\begin{equation}
    \left( \partial_z \epsilon - i\partial_z \alpha \right) = - \dfrac{\omega}{2\varepsilon_0 E_0 c} U_{xq}^{\mathrm{T}} \mathcal{P}^q
    \label{eq:observables_vs_polarization}
\end{equation}

\subsection{Hamiltonian of the system}

The quantum-mechanical evolution of a quantum state, described using the density matrix, can be calculated using the Liouville equation
\begin{equation}
     \dfrac{d}{dt}\rho = -\dfrac{i}{\hbar} \Big[ \hat{H}, \rho \Big] - \dfrac{1}{2}\Big\lbrace \hat{\Gamma}, \rho \Big\rbrace +  \hat{\Lambda} \left( \rho \right),
     \label{eq:LiouvilleGeneric}
\end{equation}
where $\hat{H}$ is the total Hamiltonian of the system and $\hat{\Gamma}$ and $\hat{\Lambda}$ are relaxation operators responsible for relaxation and repopulation of the levels (see, for example, Refs.~\cite{Auzinsh2010OpticallyInteractions,Barrat1961}).  The Hamiltonian $\hat{H}$ is the sum of the unperturbed Hamiltonian $\hat{H}_0$ and the interaction operator $\hat{V}$, which in the considered case accounts for the interaction with light and the magnetic field. In the rotating frame, the unperturbed Hamiltonian is given by $\hat{\tilde{H}}_0 = \hbar \Delta \hat{P}_e$, where $\Delta$ is the detuning of the light from the transition. In the rotating-wave approximation (RWA), the magnetic and optical parts of the Hamiltonian are given by
\begin{subequations}
    \begin{eqnarray}
        \hat{\tilde{H}}_{mag} =& \hbar \Omega_L \left( \hat{P}_g \hat{F}_z \hat{P}_g + \beta \hat{P}_e \hat{F}_z \hat{P}_e \right),\\
        \hat{\tilde{H}}_{opt} =& - \dfrac{E_0}{2} \left( \hat{P}_g \hat{d}_y \hat{P}_e  + \hat{P}_e  \hat{d}_y \hat{P}_g\right).
    \end{eqnarray}
\end{subequations}
The Larmor frequency $\Omega_L=g_f\mu_BB/\hbar$ is determined by the magnetic field $B$ with $g_f$ being the ground-state Land\'e factor, $\mu_B$ the Bohr magneton, $\beta = g_F/g_f$ the ratio between Land\'e factors of the excited and ground states, and $\hat{d}_y$ is the dipole operator for linearly polarized light along the $y$-axis. 

We assume that our system has two independent mechanisms of relaxation. The first originates from spontaneous emission, resulting in depopulation of the excited state with a relaxation rate $\Gamma$. The second comes from the interaction with environment and it is associated, e.g., with collisions with container walls, intraatomic collisions, etc. This relaxation is typically modeled as isotropic with rate $\gamma$. In turn, the relaxation operator is given by
\begin{equation}
    \hat{\Gamma} = \hat{\tilde{\Gamma}} = \hat{P}_e \Gamma + \hat{\mathbb{1}} \gamma
\end{equation}

Since we want to conserve a number of atoms in our sample, we need to introduce repopulation. In the rotating frame, the repopulation due to the spontaneous emission is given by \cite{Barrat1961}
\begin{equation}
    \hat{\Lambda}_{sp}(\tilde{\rho}) = 2 \Gamma \sum_i \hat{P}_g \left( \hat{d}_i \tilde{\rho} \hat{d}_i \right) \hat{P}_g.
\end{equation}
At the same time, isotropic relaxation should equally repopulate the entire ground state, so that
\begin{equation}
    \hat{\Lambda}_{iso} (\tilde{\rho}) = \dfrac{\gamma}{2f+1} \hat{P}_g
\end{equation}

\subsection{Evolution of optical coherences}

It can be deduced from Eqs.~\eqref{eq:complex_polarization} and \eqref{eq:observables_vs_polarization} that in order to calculate the polarization rotation and the ellipticity change, the evolution of the optical coherences needs to be determined. Here, we assume weak probe light, hence the density matrix can be expanded into the power series over $E_0$ and cut at the linear term, i.e., $\tilde{\rho} = \sum_{n} \tilde{\rho}^{(n)} E_0^n \approx \tilde{\rho}^{(0)} + \tilde{\rho}^{(1)} E_0$
\begin{widetext}
    \begin{subequations}
        \begin{eqnarray}
            \dfrac{d}{dt} \tilde{\rho}_{\mu m}^{(0)}(t) &=& - \left\{ i\left[ \Delta + \Omega_L(\beta \mu - m) \right] + \dfrac{2\gamma + \Gamma}{2} \right\} \tilde{\rho}_{\mu m}^{(0)}(t),\\
            \dfrac{d}{dt} \tilde{\rho}_{\mu m}^{(1)}(t) &=& - \left\{ i\left[ \Delta + \Omega_L(\beta \mu - m) \right] + \dfrac{2\gamma + \Gamma}{2} \right\} \tilde{\rho}_{\mu m}^{(1)}(t) - \dfrac{i}{2\hbar} \sum_n \hat{d}^y_{\mu n} \tilde{\rho}_{nm}^{(0)}(t).
            \label{eq:ground_state_evo}
        \end{eqnarray}
    \end{subequations}
\end{widetext}
This allows us to express the density-matrix evolution as
\begin{widetext}
    \begin{equation}
        \dfrac{d}{dt} \tilde{\rho}_{\mu m}(t) \approx - \left\{ i\left[ \Delta + \Omega_L(\beta \mu - m) \right] + \dfrac{2\gamma + \Gamma}{2} \right\} \tilde{\rho}_{\mu m}(t) - \dfrac{i E_0}{2\hbar} \sum_n \hat{d}^y_{\mu n} \tilde{\rho}_{nm}^{(0)}(t).
    \end{equation}
\end{widetext}

It is noteworthy that our evolution depends only on the ground state subspace in zeroth order so in the absence of the optical field. In addition, such an expression can be simplified. First, one should note that as we want to perform a non-destructive measurement and stay in the regime of low magnetic field (linear Zeeman effect), we get the condition $\Omega_L \ll \Delta$. Second, for room-temperature vapors, isotropic relaxation is typically a few orders of magnitude smaller than the spontaneous emission rate, $\gamma \ll \Gamma$. Finally, the Wigner-Eckart theorem can be used to represent the matrix elements of the dipole operator by their reduced matrix elements, in a manner similar to that used in Eq.~\eqref{eq:complex_polarization}.
\begin{widetext}
    \begin{equation}
    \small
            \dfrac{d}{dt}\tilde{\rho}_{\mu m}(t) = - \left( i \Delta + \dfrac{\Gamma}{2} \right) \tilde{\rho}_{\mu m}(t) + \dfrac{i \Omega_{R}}{2}(-1)^{j-f+3F+I-\mu} \sqrt{(2f+1)(2F+1)} \sixj{j}{f}{I}{F}{J}{1} \left[ \sum_{q,n} U_{yq}^{T} \threej{F}{1}{f}{-\mu}{q}{n} \rho_{nm}^{(0)}(t) \right],
    \end{equation}
\end{widetext}
where $\Omega_R = E_0 \expval{j \norm{d^{(J)}} J}/\hbar$ is the Rabi frequency. By substituting the above relation into Eqs.~\eqref{eq:complex_polarization} and \eqref{eq:observables_vs_polarization}, one can get the relations for the temporal evolution of the polarization rotation and ellipticity change
\begin{widetext}
    \begin{equation}
        \dfrac{d}{dt} \left( \partial_z \epsilon - i\partial_z \alpha \right) = - \left( i \Delta + \dfrac{\Gamma}{2} \right)\left( \partial_z \epsilon - i\partial_z \alpha \right) + i \chi \left[ \sum_{q r} U_{xq}^{\mathrm{T}}U_{yr}^{\mathrm{T}} \sum_{m n \mu} (-1)^q \threej{f}{1}{F}{-m}{q}{\mu} \threej{F}{1}{f}{-\mu}{r}{n} \tilde{\rho}_{nm}^{(0)}(t)\right],
    \label{eq:evolution_elipticity_rotation}
    \end{equation}
\end{widetext}
where $\chi$ is given by
\begin{widetext}
    \begin{equation}
        \chi = \dfrac{N_{at}\omega \abs{\expval{j  \norm{\widehat{\mathbf{d}}^{(J)}} J }}^{2} }{2 \varepsilon_0 c \hbar} (-1)^{2j+2J}(2f+1)(2F+1) \sixj{j}{f}{I}{F}{J}{1}^2\,.
    \end{equation}
\end{widetext}

\subsection{Effective observables}

Analyzing Eq.~\eqref{eq:evolution_elipticity_rotation} one can see that the evolution of the polarization rotation and change of ellipticity are related and depend only on the ground state in the zeroth-order approximation (we assumed that the optical field is very week, so that the excited-state population is negligible.) Moreover, by the analysis of Eq.~\eqref{eq:RWA} one can show that RWA does not change the ground-state subspace (since now on we drop tilde over the ground-state elements). Thus, the evolution of those ground-state terms are given by
\begin{subequations}
\begin{eqnarray}
    \tilde{\rho}_{nm}^{(0)}(t) &=& \rho_{nm}^{(0)}(0)\mathrm{e}^{-\left[\gamma -i(n-m)\right]t}\\
    \tilde{\rho}_{nn}^{(0)}(t) &=& \left( \rho_{nn}^{(0)}(0)-\dfrac{1}{2f+1}\right)\mathrm{e}^{-\gamma t} + \dfrac{1}{2f+1}
    \label{eq:ground_state_evo}
\end{eqnarray}
\end{subequations}

At time $t=0$ we begin to probe the atoms with light. Thus, at $t=0$, the zeroth-order density matrix is exactly the density matrix of the initial state, $\rho_{nm}^{(0)}(0) = \rho_{nm}(0)$. Applying this into Eq.~\eqref{eq:evolution_elipticity_rotation} and carefully collecting terms allows us to write this equation in the following form
\begin{widetext}
    \begin{equation}
         \dfrac{d}{dt}\left[ \partial_z \epsilon  - i \partial_z \alpha \right] = - \left( i \Delta + \dfrac{\Gamma}{2} \right) \left[ \partial_z \epsilon  - i \partial_z \alpha \right] + \dfrac{i \chi}{2} e^{-\gamma t} \left[ \expval{\hat{\alpha}_{R}} \sin (2 \Omega_L t) + \expval{\hat{\alpha}_{I}} \cos (2 \Omega_L t) - i \expval{\hat{\beta}} \right],
         \label{eq:ODE_elipticity_rotation}
    \end{equation}
\end{widetext}
where we introduced three observables
\begin{widetext}
    \begin{subequations}
        \begin{eqnarray}
            \hat{\alpha}_{R} &=& \sum_{n=-f}^{f-2} \threej{f}{1}{F}{-n-2}{1}{n+1} \threej{F}{1}{f}{-n-1}{1}{n} \left( \dyad{n}{n+2} + \dyad{n+2}{n} \right),\\
            \hat{\alpha}_{I} &=& \sum_{n=-f}^{f-2} \threej{f}{1}{F}{-n-2}{1}{n+1} \threej{F}{1}{f}{-n-1}{1}{n} \left( i\dyad{n}{n+2} - i\dyad{n+2}{n} \right),\\
            \hat{\beta} &=& \sum_{n=-f}^{f} \left[ \threej{f}{1}{F}{-n}{1}{n-1}^{2} - \threej{f}{1}{F}{-n}{-1}{n+1}^{2} \right] \dyad{n}.
        \end{eqnarray}
        \label{eq:Observables}
    \end{subequations}
\end{widetext}

We separate the real and imaginary parts of Eq.~\eqref{eq:ODE_elipticity_rotation}, which allows us to solve the equations analytically. Although general solutions of the equations are complicated, they can be simplified using the assumptions discussed above. However, it can be significantly simplified using the following assumptions. First, it should be noted that we are mostly interested in the evolution connected with the drive originating from the ground state evolution, this evolution takes place on the time scale characterized by the ground state decay rate $\gamma$ and Larmor frequency $\Omega_L$. Thus, we can neglect the transient effects proportional to $\exp\left( -\Gamma t \right)$. Next, we can group the oscillatory and constant terms of the remaining solution and expand the respective amplitudes in the power series to the first order in $\Omega_L$ and $\gamma$. The mentioned amplitudes are the functions of detuning $\Delta$, Larmor frequency $\Omega_L$ and relaxation rates $\gamma$ and $\Gamma$. As under typical experimental conditions \cite{Pustelny2006Pump-probeLight, Pustelny2011Tailoring}, $\Omega_L$ and $\gamma$ span between several Hz up to tens of kHz. At the same time, the relaxation rate of the first excited state of typically investigated alkali atoms is in the MHz range, so that to fulfill the assumptions the detuning should be at least several times larger. Finally, under typical conditions, the room temperature alkali vapors are optically thin, so that $\partial_z \alpha$ can be written in linear limit $\Delta \alpha / L$, where $L$ is the thickness of the atomic medium. Altogether, this allows us to achieve quite a compact form of the expressions 
\begin{widetext}
    \begin{subequations}
        \begin{eqnarray}
            \Delta \alpha =& - \chi L \mathrm{e}^{-\gamma t } \left( \dfrac{\Gamma \expval{\hat{\alpha}_r}}{\Gamma^2 + 4 \Delta^2}\sin \left( 2 \Omega_L t \right) + \dfrac{\Gamma \expval{\hat{\alpha}_i}}{\Gamma^2 + 4 \Delta^2} \cos \left( 2 \Omega_L t \right) - \dfrac{2 \Delta \expval{\hat{\beta}}}{\Gamma^2 + 4 \Delta^2}  \right)\,,\\
            \Delta \epsilon =&  \chi L \mathrm{e}^{-\gamma t } \left( \dfrac{2 \Delta \expval{\hat{\alpha}_r}}{\Gamma^2 + 4 \Delta^2}\sin \left( 2 \Omega_L t \right) + \dfrac{2 \Delta \expval{\hat{\alpha}_i}}{\Gamma^2 + 4 \Delta^2} \cos \left( 2 \Omega_L t \right) + \dfrac{\Gamma \expval{\hat{\beta}}}{\Gamma^2 + 4 \Delta^2}  \right)\,.
        \end{eqnarray}
    \end{subequations}
\end{widetext}
It should be noted that $\Gamma / \left( \Gamma^2 +4 \Delta^2 \right)$ and $2\Delta / \left( \Gamma ^2 + 4 \Delta^2 \right)$ are, in fact, the real and imaginary part of the Lorentz function $\mathcal{L}(\Delta) = (\Gamma - 2i \Delta)^{-1}$.

In analogy to the polarization rotation and ellipticity change, relations for absorption and phase change can be derived. To do so we introduce another observable 
\begin{equation}
    \hat{\delta} = \sum_{n=-f}^{f} \left[ \threej{f}{1}{F}{-n}{1}{n-1}^{2} + \threej{f}{1}{F}{-n}{-1}{n+1}^{2} \right] \dyad{n}.
\end{equation}
which is used in the following relations for absorption $\Delta E$ and phase change $\Delta\varphi$
\begin{widetext}
    \begin{subequations}
        \begin{eqnarray}
            \dfrac{\Delta E}{E} &=& \chi L \mathcal{L}_R(\Delta) \mathrm{e}^{-\gamma t } \left( \expval{\hat{\alpha}_r} \cos \left( 2 \Omega_L t \right) - \expval{\hat{\alpha}_i} \sin \left( 2 \Omega_L t \right) + \left(\expval{\hat{\delta}}-\delta_{s}\right) \right) + \chi L \mathcal{L}_R (\Delta) \delta_{s}\,,\\
            \Delta \varphi &=&  - \chi L \mathcal{L}_I(\Delta) \mathrm{e}^{-\gamma t } \left( \expval{\hat{\alpha}_r} \cos \left( 2 \Omega_L t \right) - \expval{\hat{\alpha}_i} \sin \left( 2 \Omega_L t \right) + \left(\expval{\hat{\delta}}-\delta_{s}\right) \right) - \chi L \mathcal{L}_I (\Delta) \delta_{s}\,.    
        \end{eqnarray}
    \end{subequations}
\end{widetext}

It should be noted that the term $\delta_{s} = \dfrac{2}{2f+1} \sum_{n=-f}^{f} \threej{f}{1}{F}{-n}{1}{n-1}^{2}$, which arises exclusively in the relations of absorption and phase change, corresponds to the isotropic fraction of the state. This is intuitively understood, as absorption and phase changes are always present in the medium, independently of its anisotropic.

\section{Appendix B: Reconstruction of specific states}

Although the main text presents a reconstruction of two generic states, i.e., pure and partially-mixed states (some analysis were also performed for the fully mixed state), here we present results of the reconstructions of several other examples that are often experimentally generated \cite{Gawlik2006NonlinearLight, Pustelny2006Pump-probeLight, Katz2018LightVapor}.

The simplest state one may consider for reconstruction is a fully mixed state, corresponding to atoms in thermal equilibrium in the absence of the magnetic field. In such a case, population of all ground-state sublevels is equal and no coherences are present in the system. It can be shown, using the angular momentum probability surfaces (AMPS) \cite{Rochester2001Atomic},which conveniently represent the density matrices as 3D objects, that the thermal-equilibrium state corresponds is fully symmetric (AMPS is a sphere). In turn, the state is optically isotropic, i.e., no polarization rotation or ellipticity change is expected. 

\begin{widetext}

\begin{figure}[h]
    \begin{tabular}{ccc}
        \includegraphics[width=0.3\textwidth]{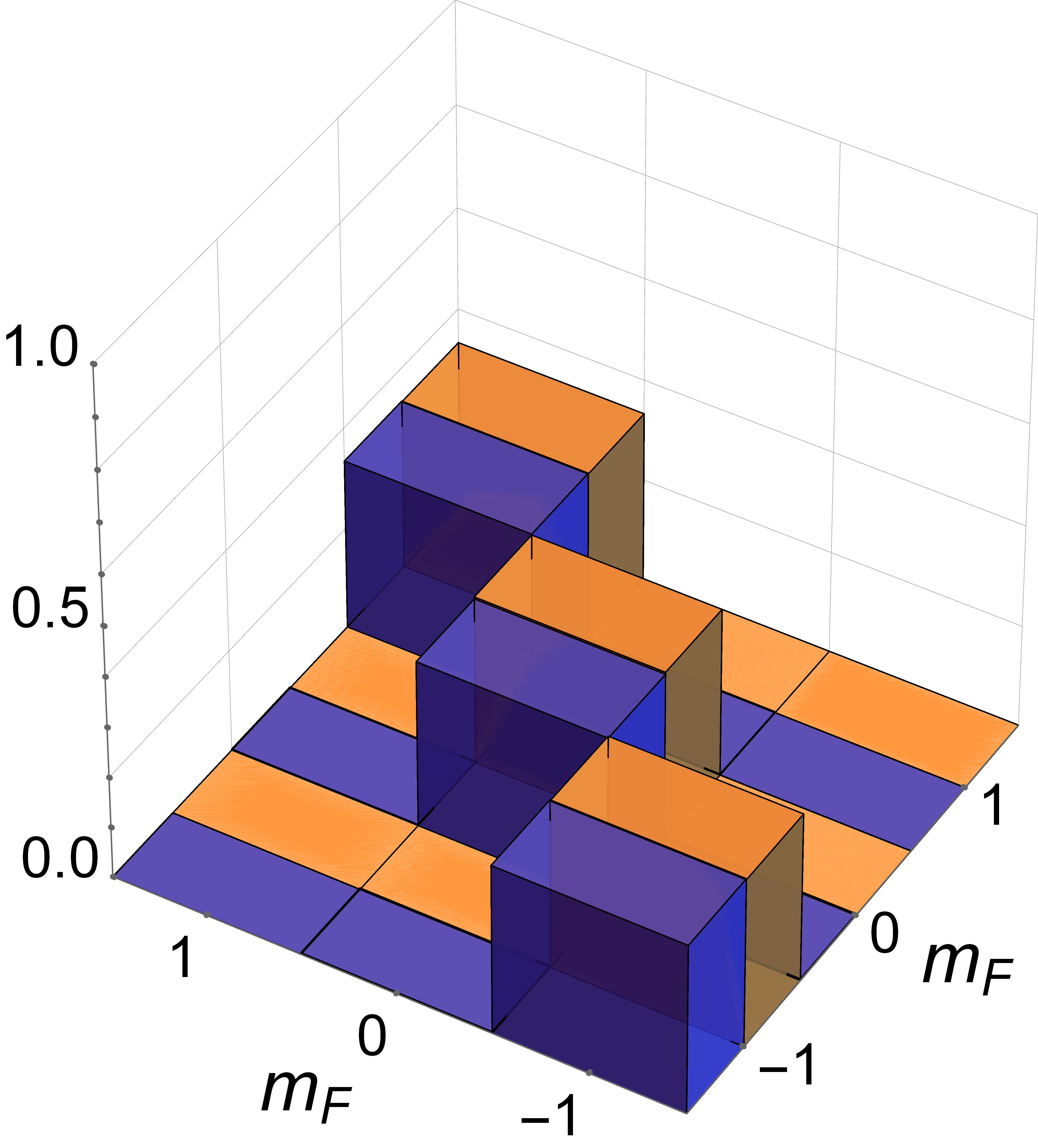} &  
        \includegraphics[width=0.3\textwidth]{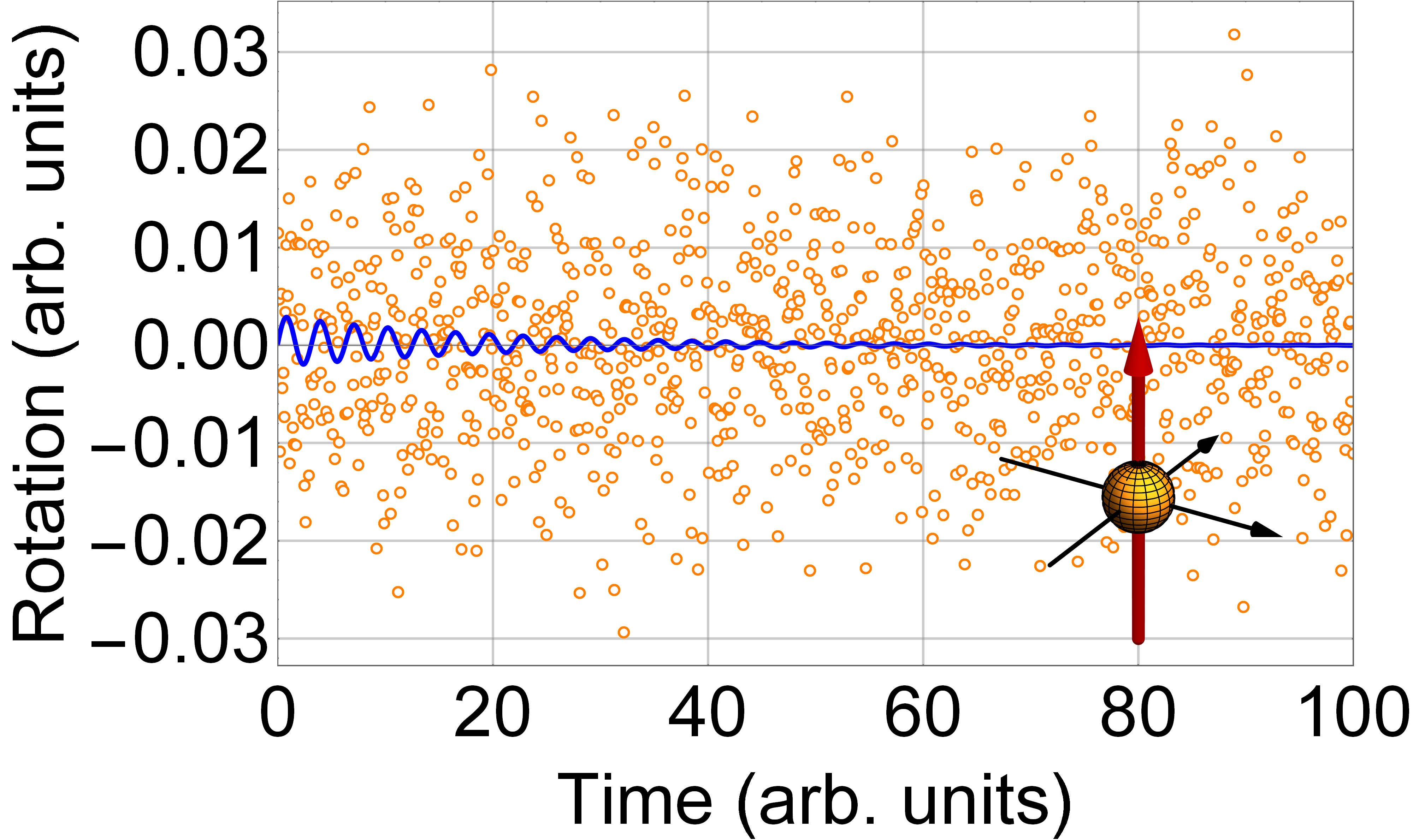} &   \includegraphics[width=0.3\textwidth]{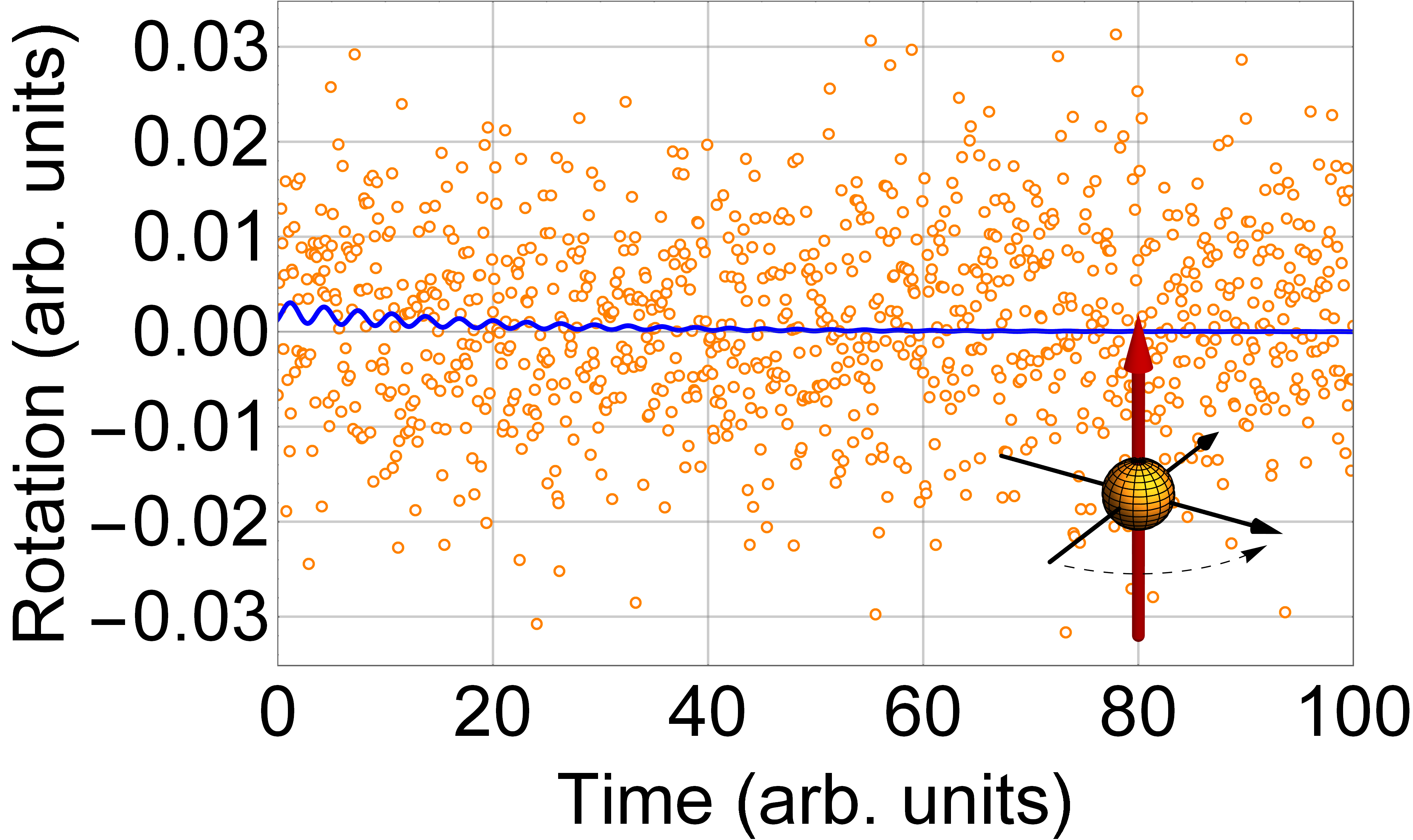} \\
        a) & c) & e) \\[6pt]
        \includegraphics[width=0.3\textwidth]{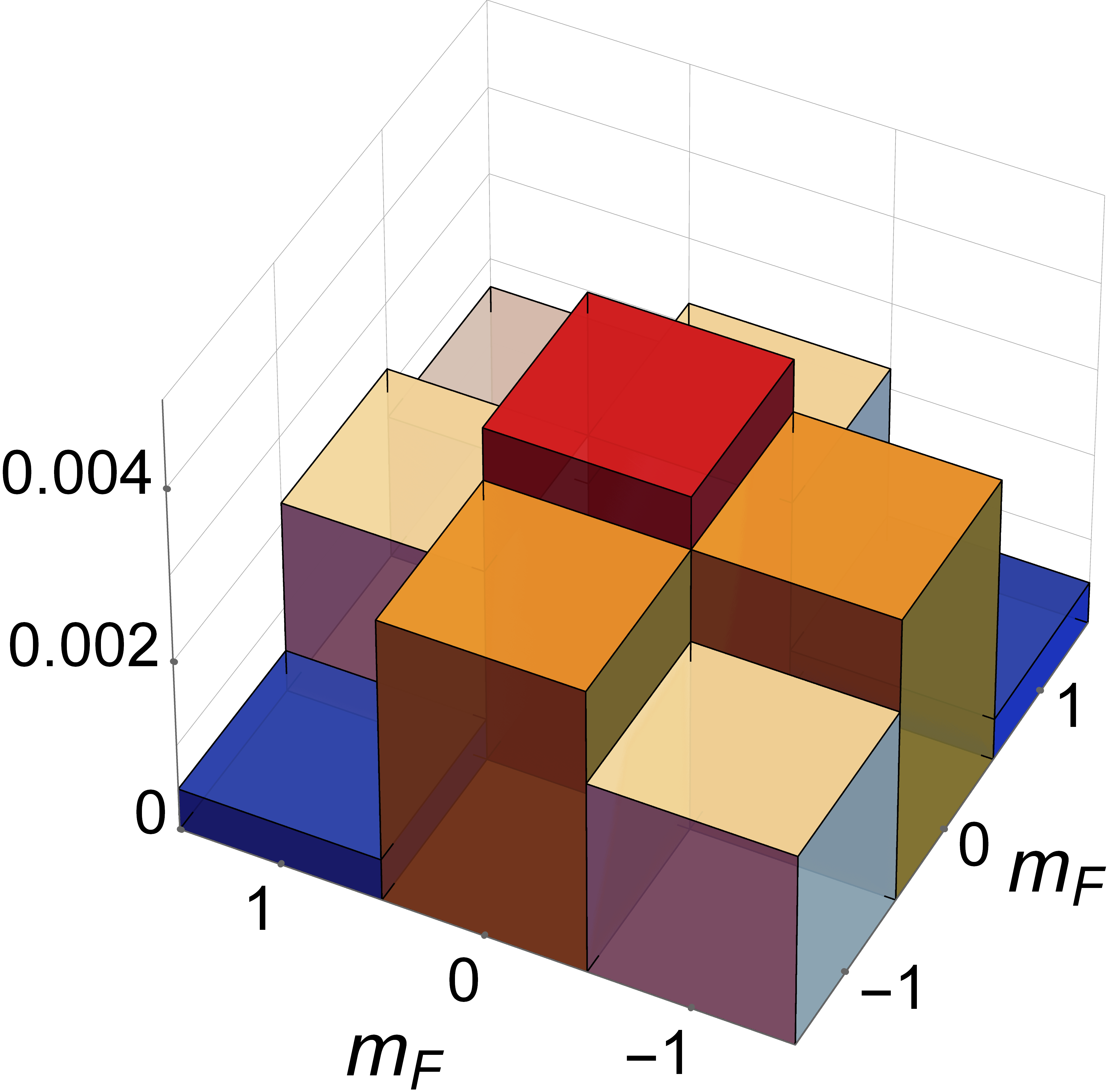} &
        \includegraphics[width=0.3\textwidth]{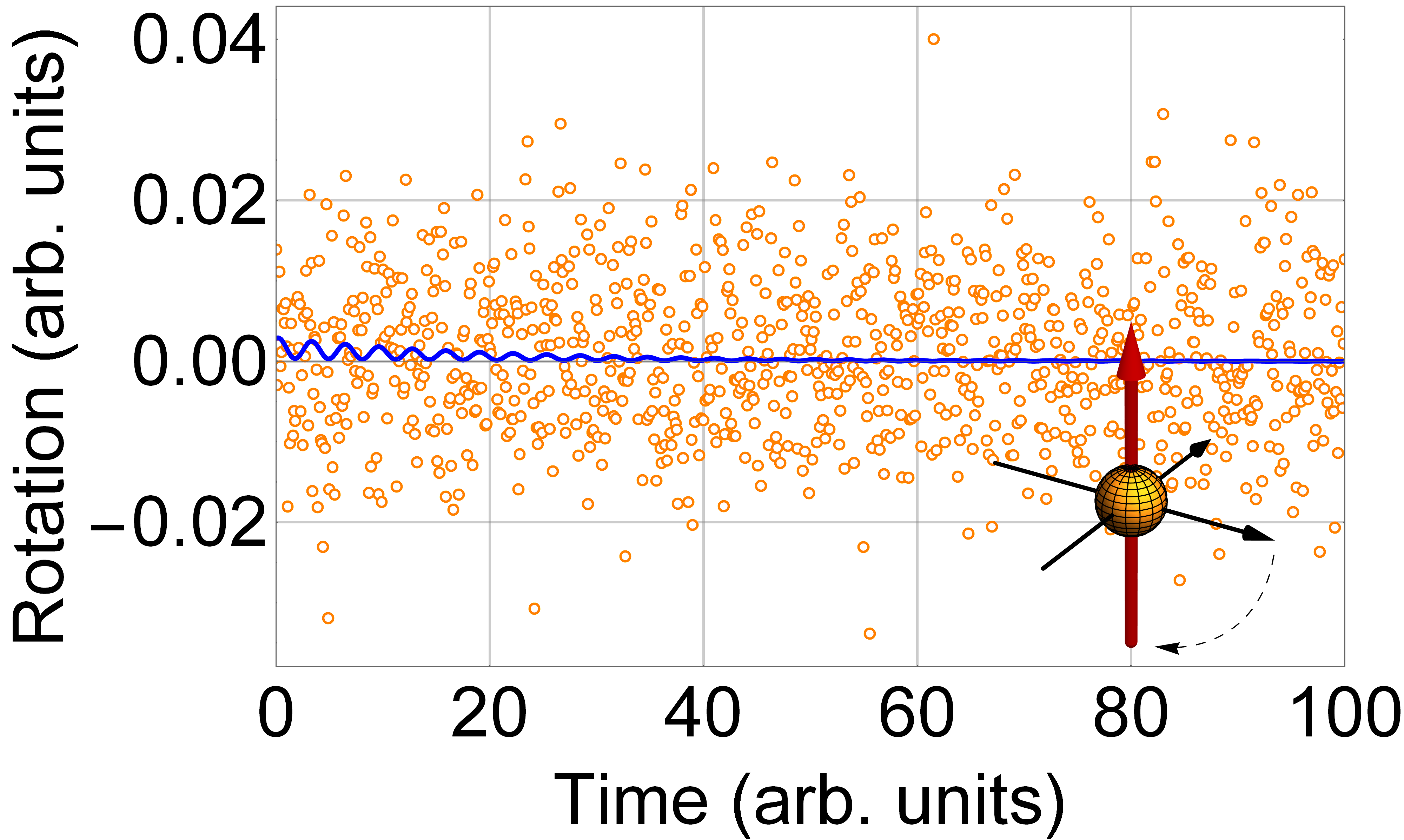} &   \includegraphics[width=0.3\textwidth]{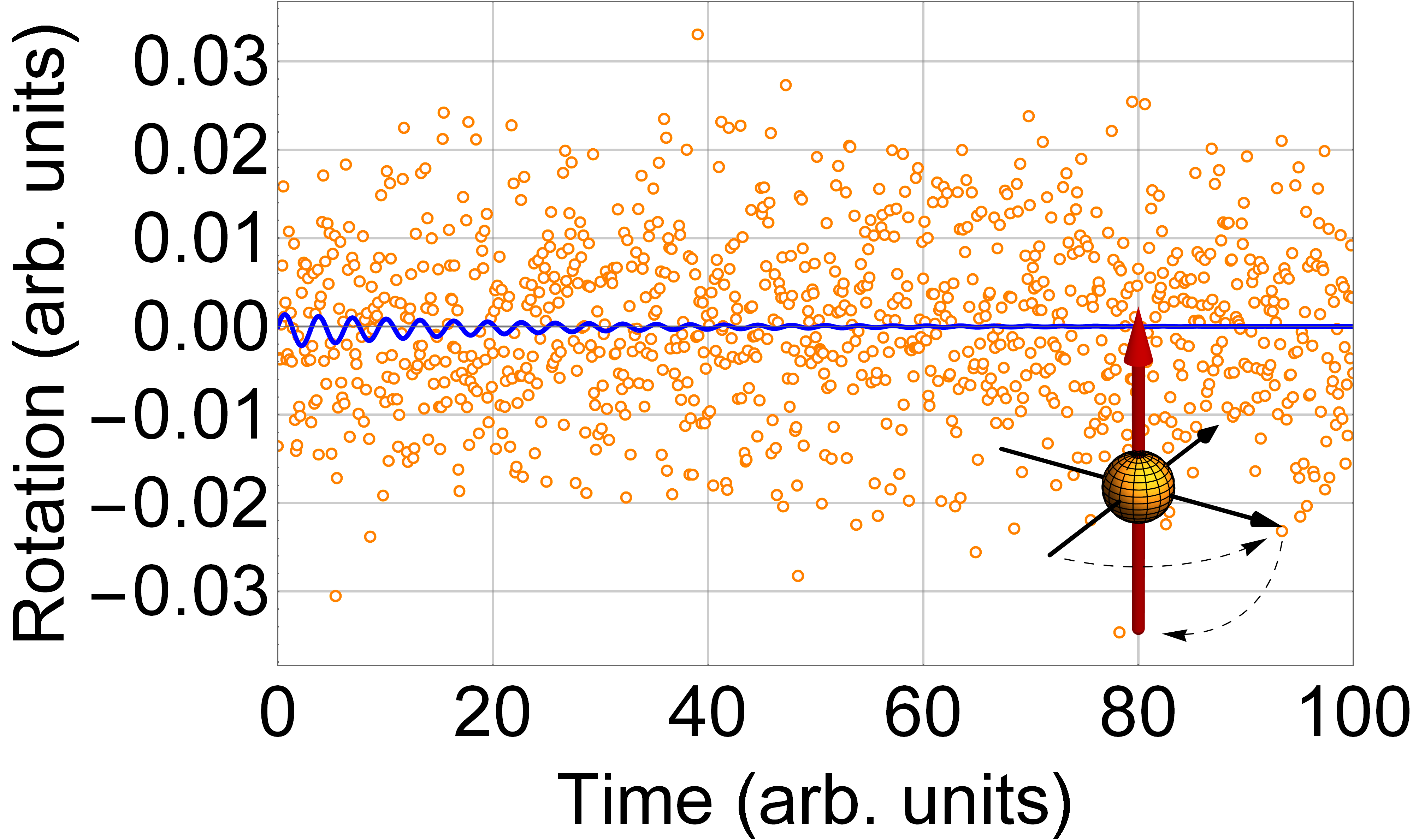} \\
        b) &  d) & f) \\[6pt]
    \end{tabular}
    \caption{Reconstruction of the fully mixed state. a) Absolute values of density-matrix elements corresponding to initial (orange/light grey) and reconstructed (blue/dark grey) states and b) the absolute differences between the corresponding elements of the density matrices. The rotation signals (orange points) along with the fitting (blue line) for c) $\varphi=0$ and $\theta=0$, d) $\varphi=0$ and $\theta=\pi/4$, e) $\varphi=\pi/4$ and $\theta=0$, and f) $\varphi=\pi/4$ and $\theta=\pi/4$. Insets in c)-f) show the AMPS of the state after respective rotations. Red thicker arrow denotes the magnetic-field/light-propagation direction and dashed arrows indicates direction of the rotation due to the control pulses. Simulation parameters identical as in the main text.}
    \label{fig:IsotropicState}
\end{figure}

\end{widetext}

We also expect this because of the full symmetry of the density matrix. It should be noted, however, that in the real measurement the signal is contaminated with noise or systematic effects, associated with imperfections of detection, which can lead to fitting of nonzero polarization rotation.

The results of the simulations of the polarization rotation proceeded with pulses rotating density matrix by the angles $\varphi$ and $\theta$ are shown in Fig.~\ref{fig:IsotropicState}c)-f).  To better simulate the experimental signals, the simulated data was contaminated with white noise, and such signals were used for fitting and reconstruction (blue traces in the experimental data).  Figure~\ref{fig:IsotropicState}a) shows initial and reconstructed density matrix and Fig.~\ref{fig:IsotropicState}b) presents deviation of the reconstructed matrix from the initial matrix. Under such conditions, the fidelity of the reconstruction is very close to unity, $\mathcal{F}=0.999$. At this point, it is hard to determine whether this remaining deviation is a consequence of the principal uncertainty of our method resulting from the approximations in the model or fitting errors.

\begin{widetext}

\begin{figure}
    \begin{tabular}{ccc}
        \includegraphics[width=0.3\textwidth]{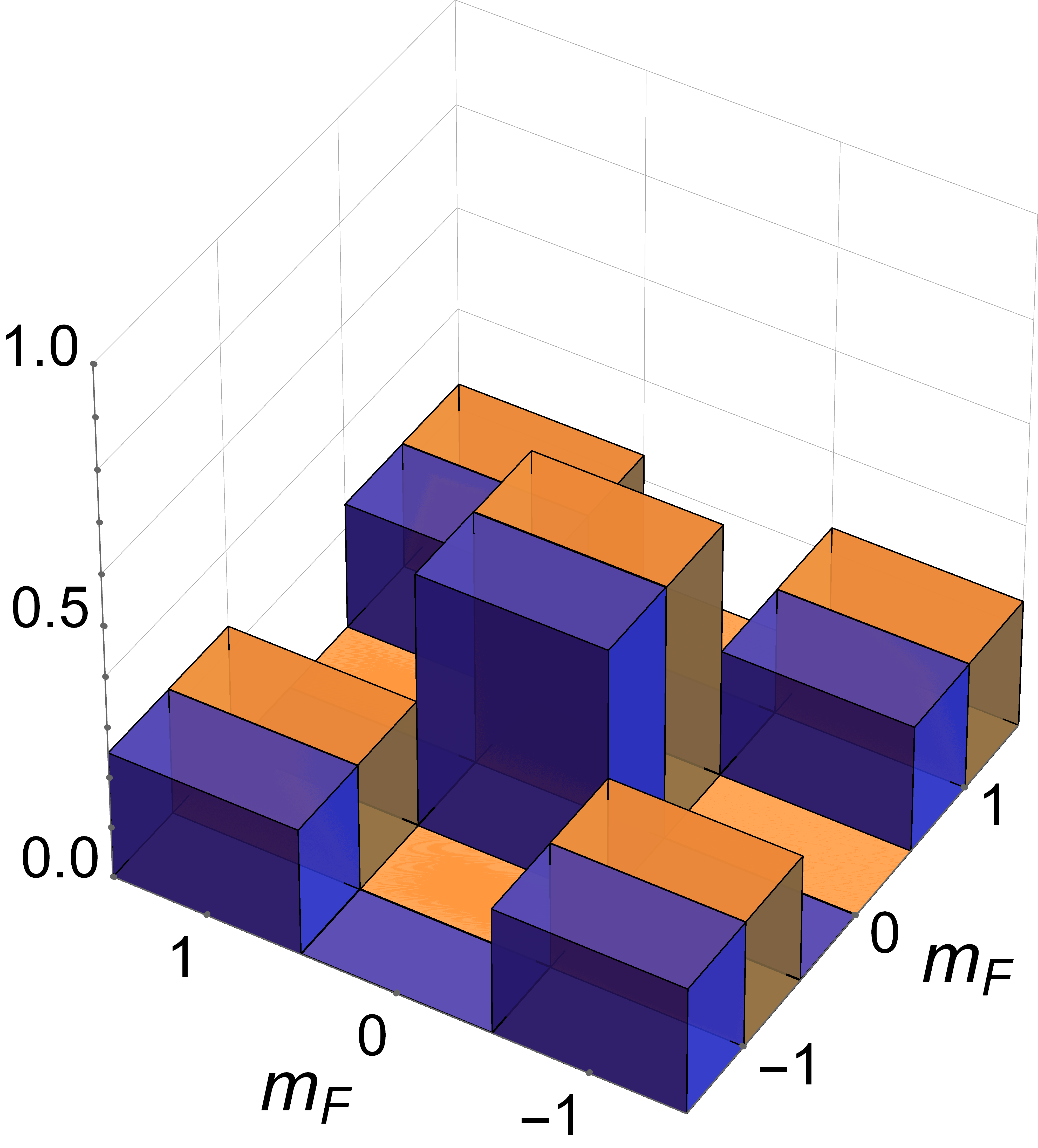} &  
        \includegraphics[width=0.3\textwidth]{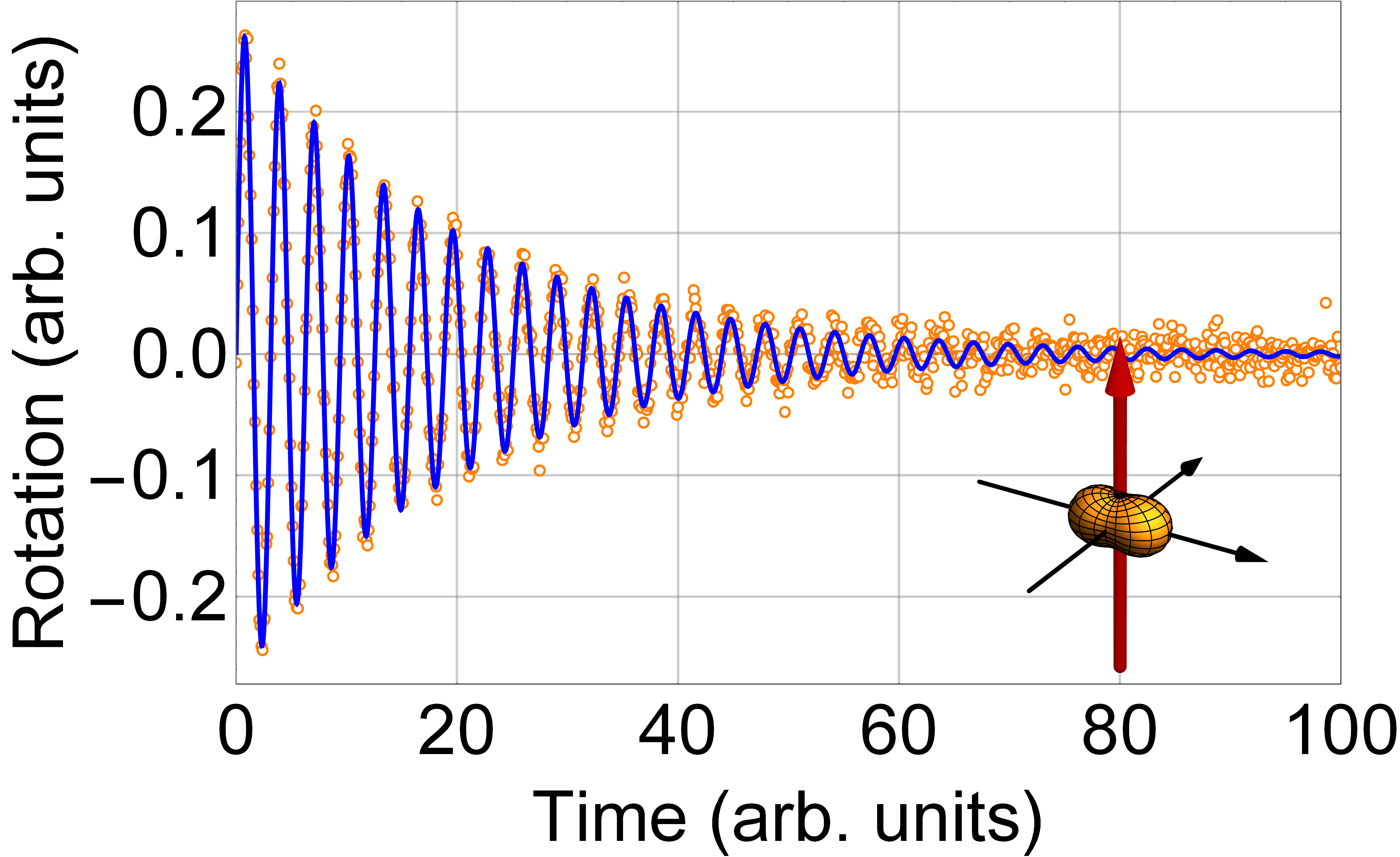} &   \includegraphics[width=0.3\textwidth]{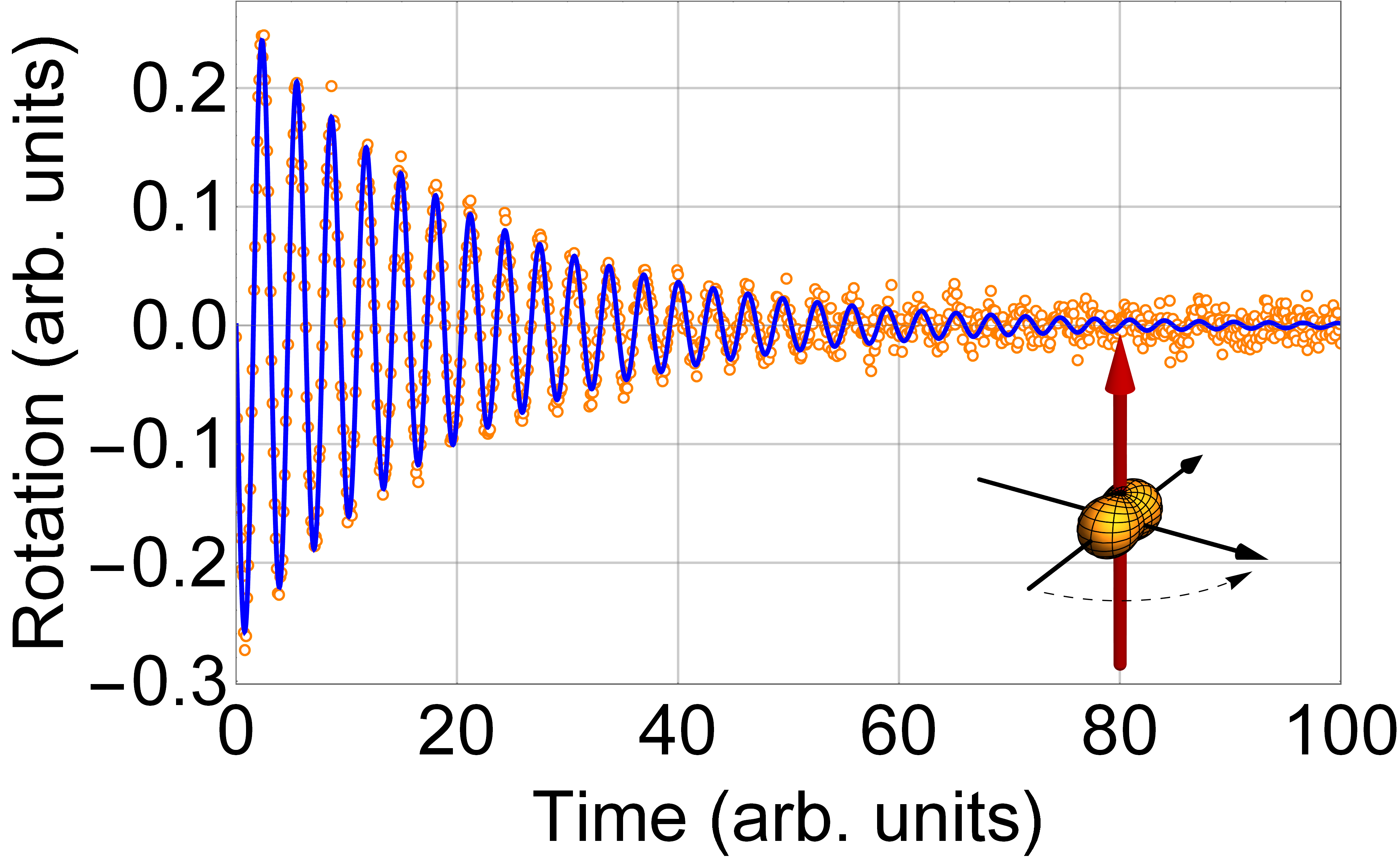} \\
        a) & c) & e) \\[6pt]
        \includegraphics[width=0.3\textwidth]{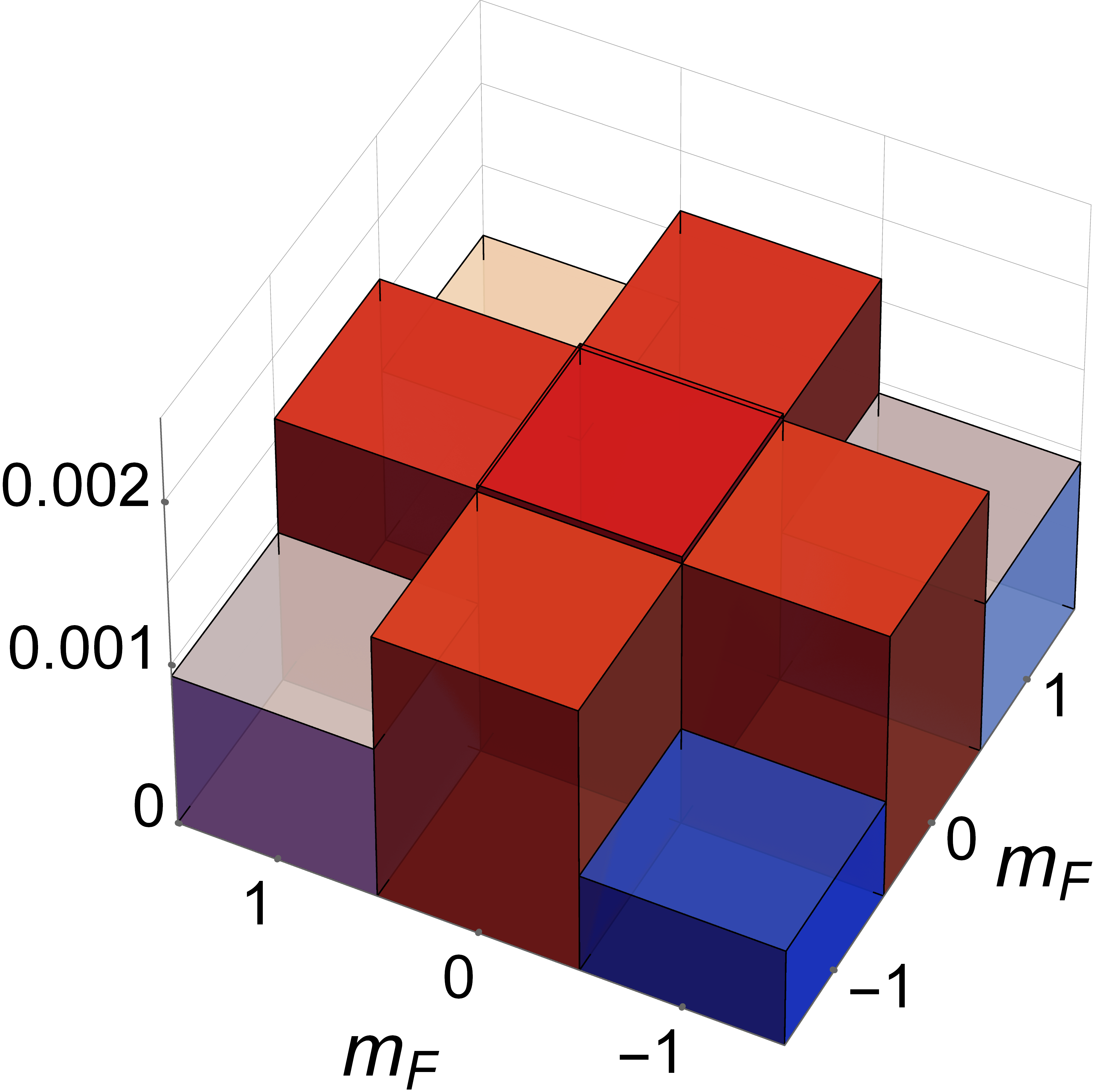} &
        \includegraphics[width=0.3\textwidth]{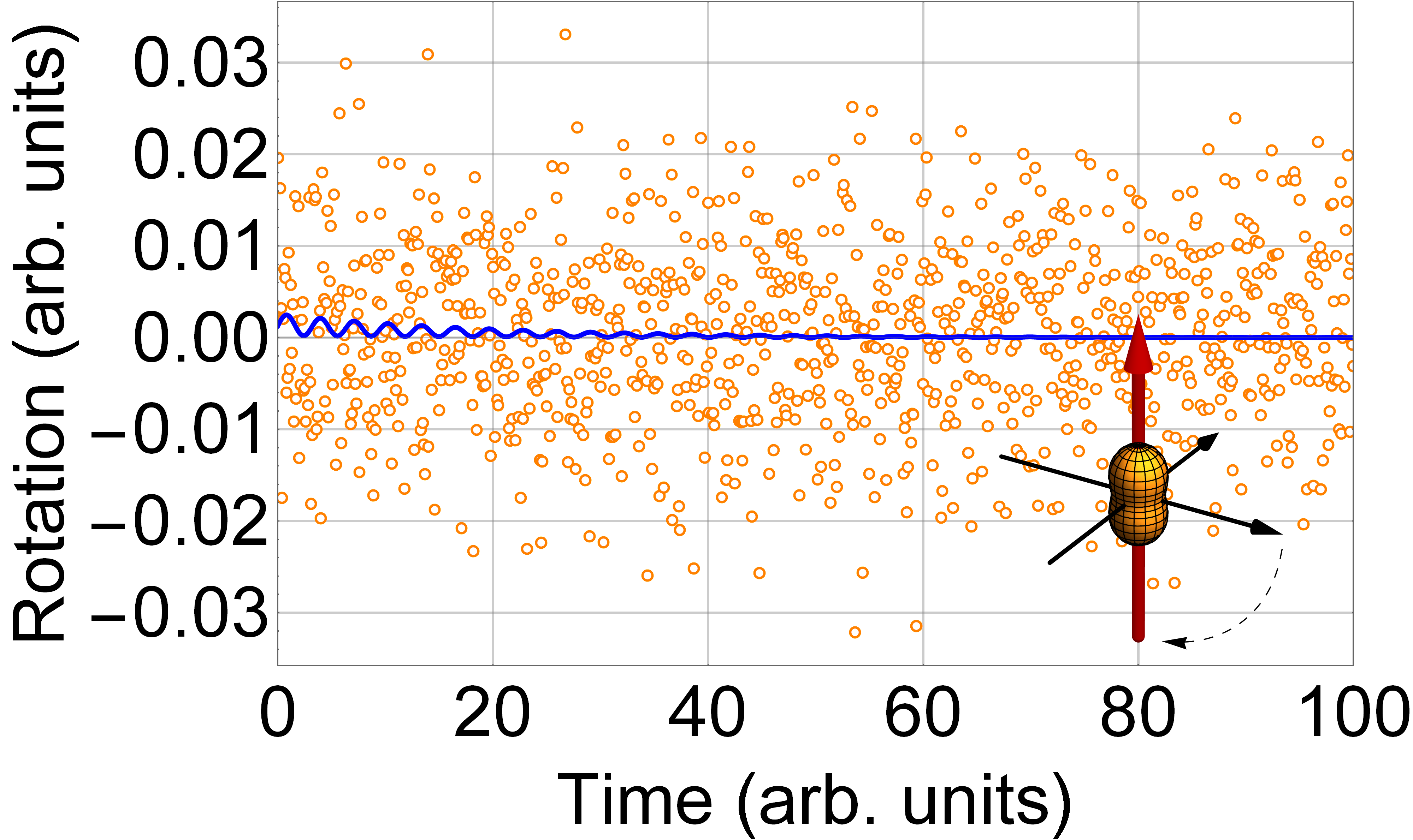} &   \includegraphics[width=0.3\textwidth]{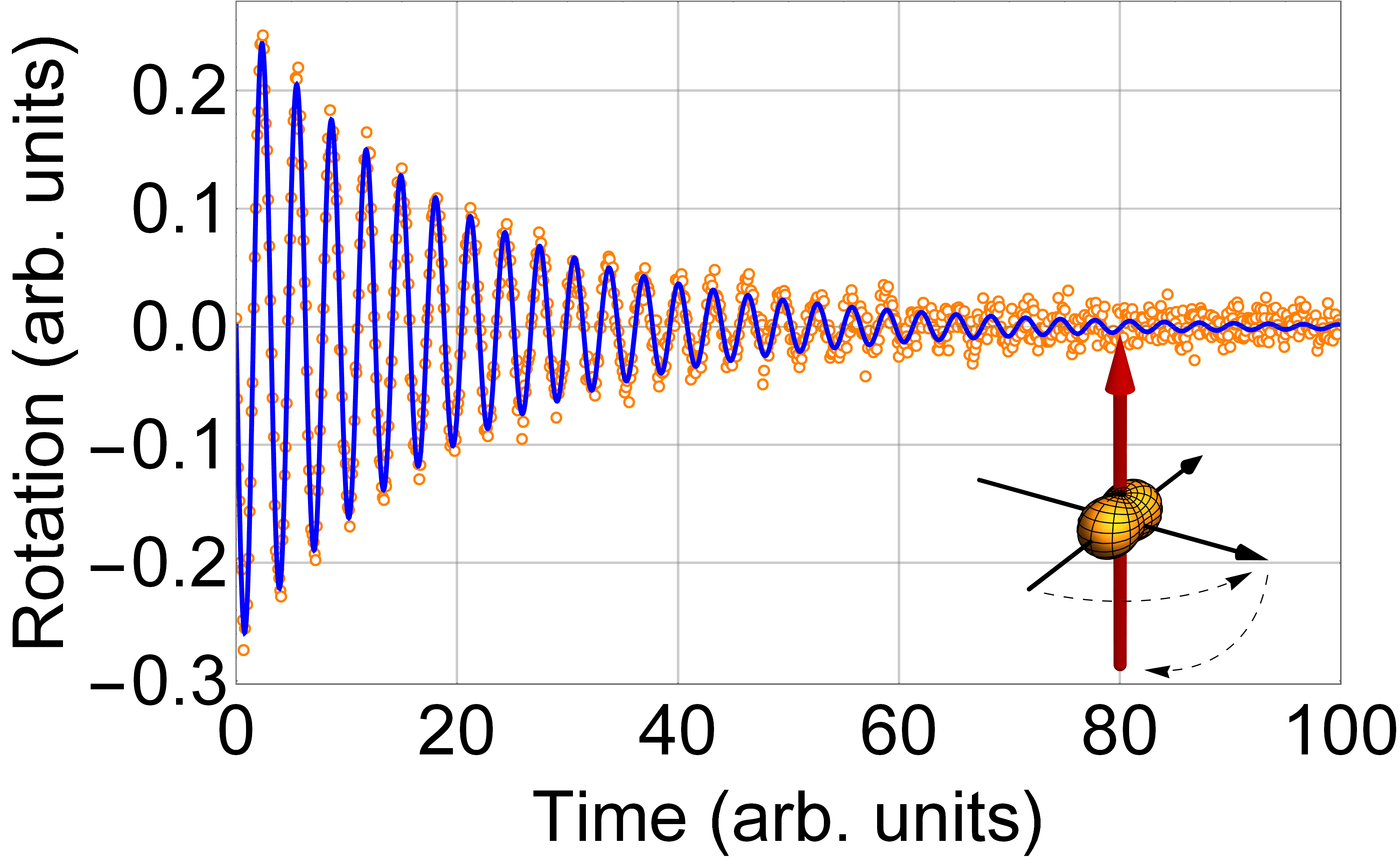} \\
        b) & d) & f) \\[6pt]
    \end{tabular}
    \caption{Reconstruction of the density matrix of a fully aligned state. a) Absolute values of the density-matrix elements corresponding to initial (orange/light grey) and reconstructed (blue/dark grey) states with b) the absolute value of the differences between the two matrices. c)-f) Simulated signals obtained for the same set of parameters as in Fig.~\ref{fig:IsotropicState}. Insets in c)-f) show the AMPS of the state after respective rotations.}
    \label{fig:AlignmentState}
\end{figure}

\end{widetext}

An aligned state is a nontrivial state that is generated, for example, by linearly polarized light.  Such a state is widely used in optical magnetometry \cite{Pustelny2008Magnetometry,Chalupczak2013Radio,Ding2016Sensitivity,Wilson2020Wide,Put2019DifferentArrangement}. When generated with $x$- or $y$-polarized light and without any transverse magnetic fields, a characteristic feature of the state is the presence of strong $\Delta m =2$ coherences with the simultaneous absence of $\Delta m =1$ coherences.  The density matrix corresponding to such a state is shown in Fig.~\ref{fig:AlignmentState}a) with an initial state shown in orange and the reconstructed state in blue.  Based on the geometric argument using AMPS (the density matrix corresponds to the peanut-like shape oriented along the polarization direction) evolution of the matrix (rotation of the peanut-like shape) due to the longitudinal magnetic field leads to the modulation of light parameters at twice the Larmor frequency (the matrix reproduces itself twice during the Larmor period). At the same time, rotation and differently oriented fields give signals in the first and second harmonics of the Larmor frequency \cite{Pustelny2006Nonlinear}, with the distinct exception of rotation around the peanut axis, which provides no modulation (full symmetry around the axis) (Fig.~\ref{fig:AlignmentState}d). The simulated signals (with noise added) along with the fitting functions are shown in Fig.~\ref{fig:AlignmentState}c)-f).
Based on the signals, we performed the reconstruction and achieved a fidelity of 0.998.  

\begin{widetext}

\begin{figure}[h]
     \begin{tabular}{ccc}
        \includegraphics[width=0.3\textwidth]{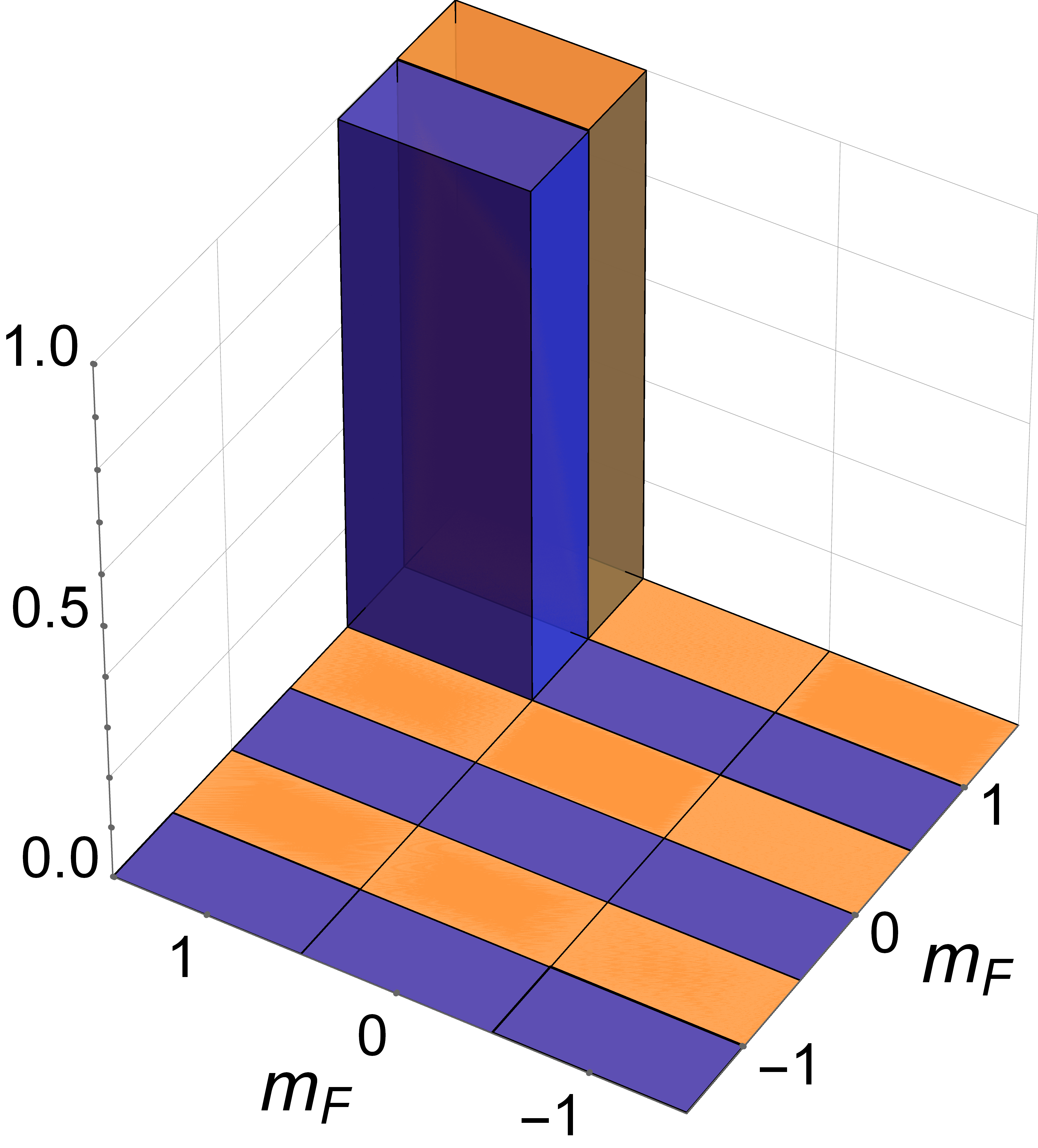} &  
        \includegraphics[width=0.3\textwidth]{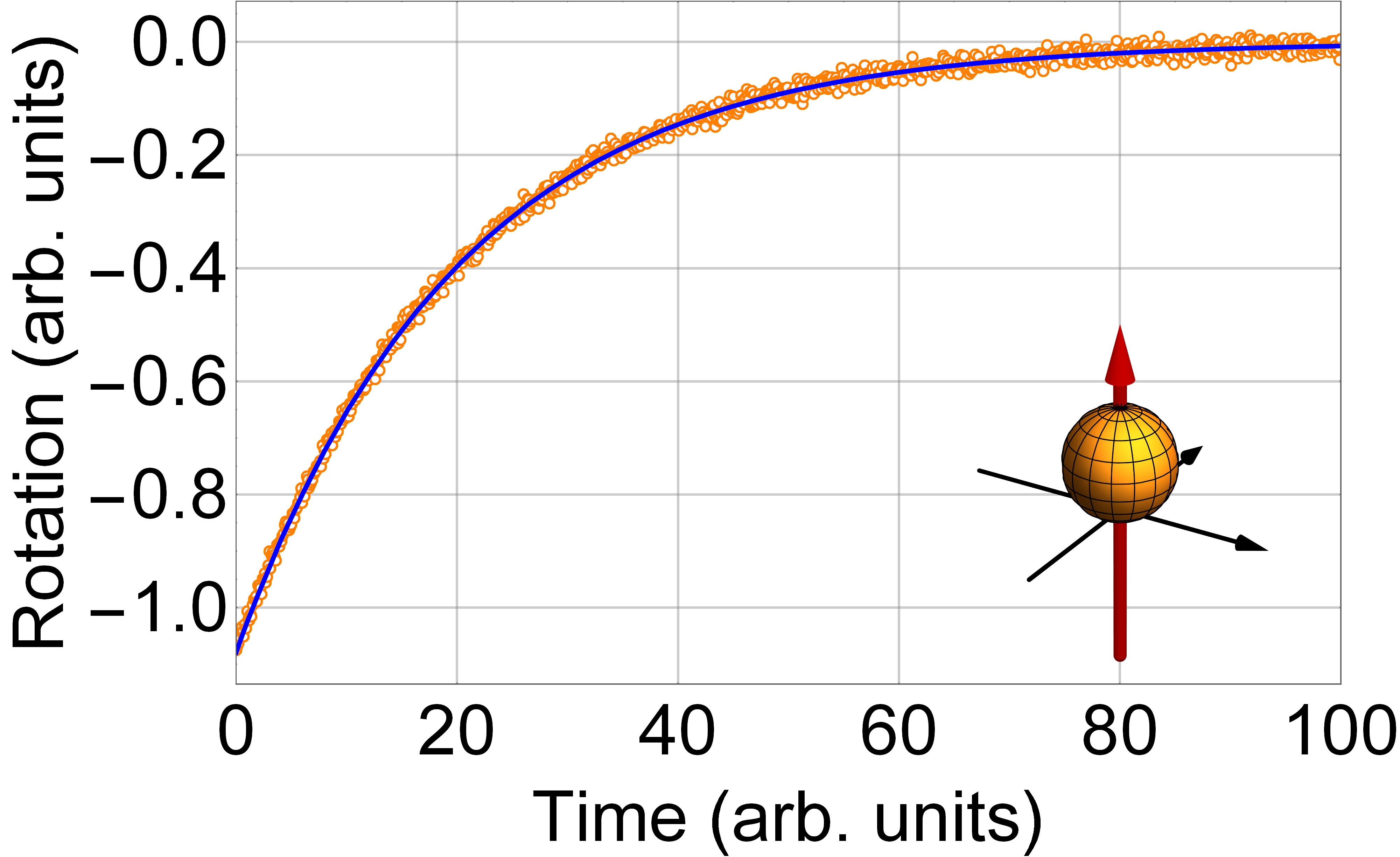} &   \includegraphics[width=0.3\textwidth]{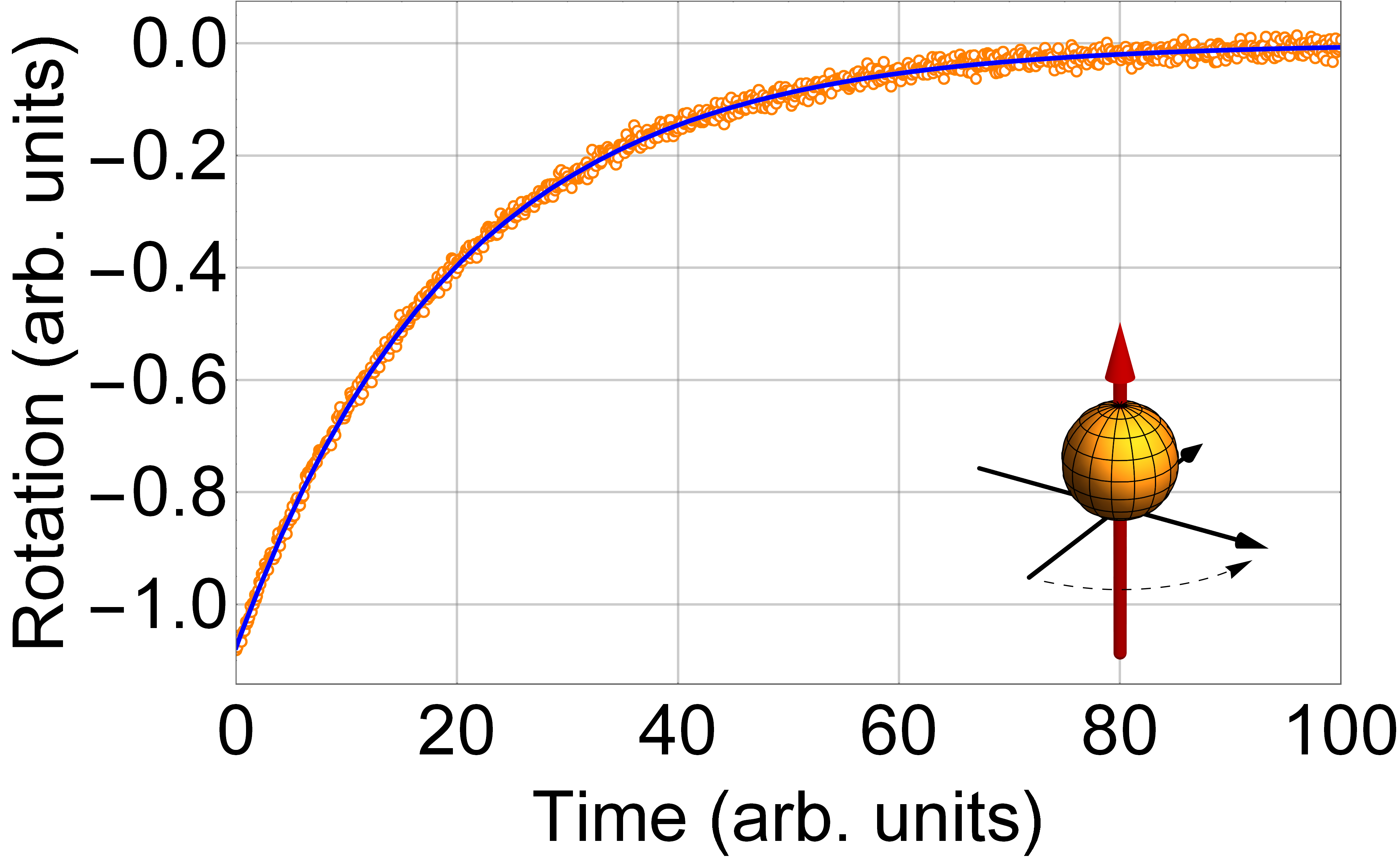} \\
        a) & c) &  e) \\[6pt]
        \includegraphics[width=0.3\textwidth]{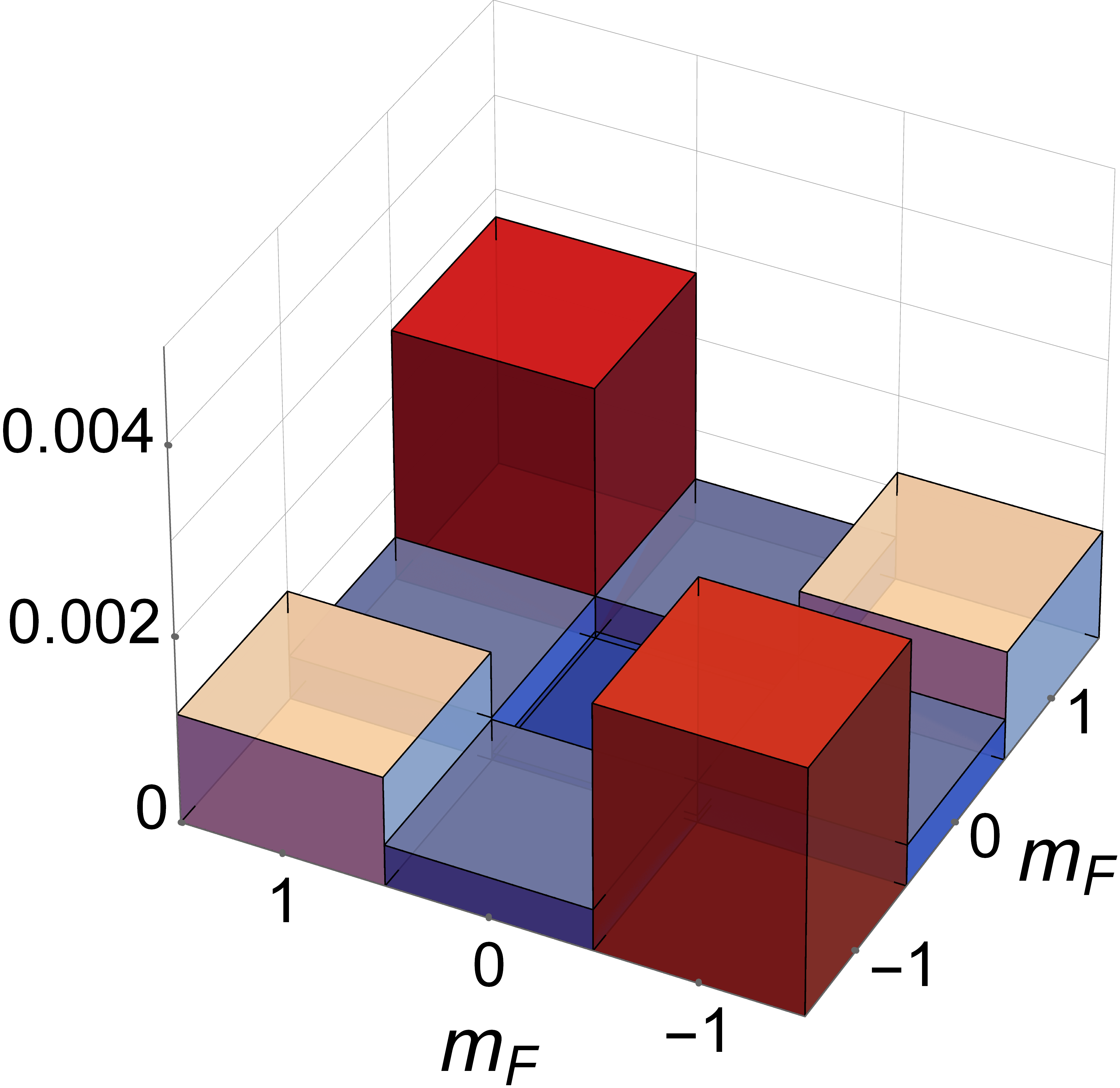} &
        \includegraphics[width=0.3\textwidth]{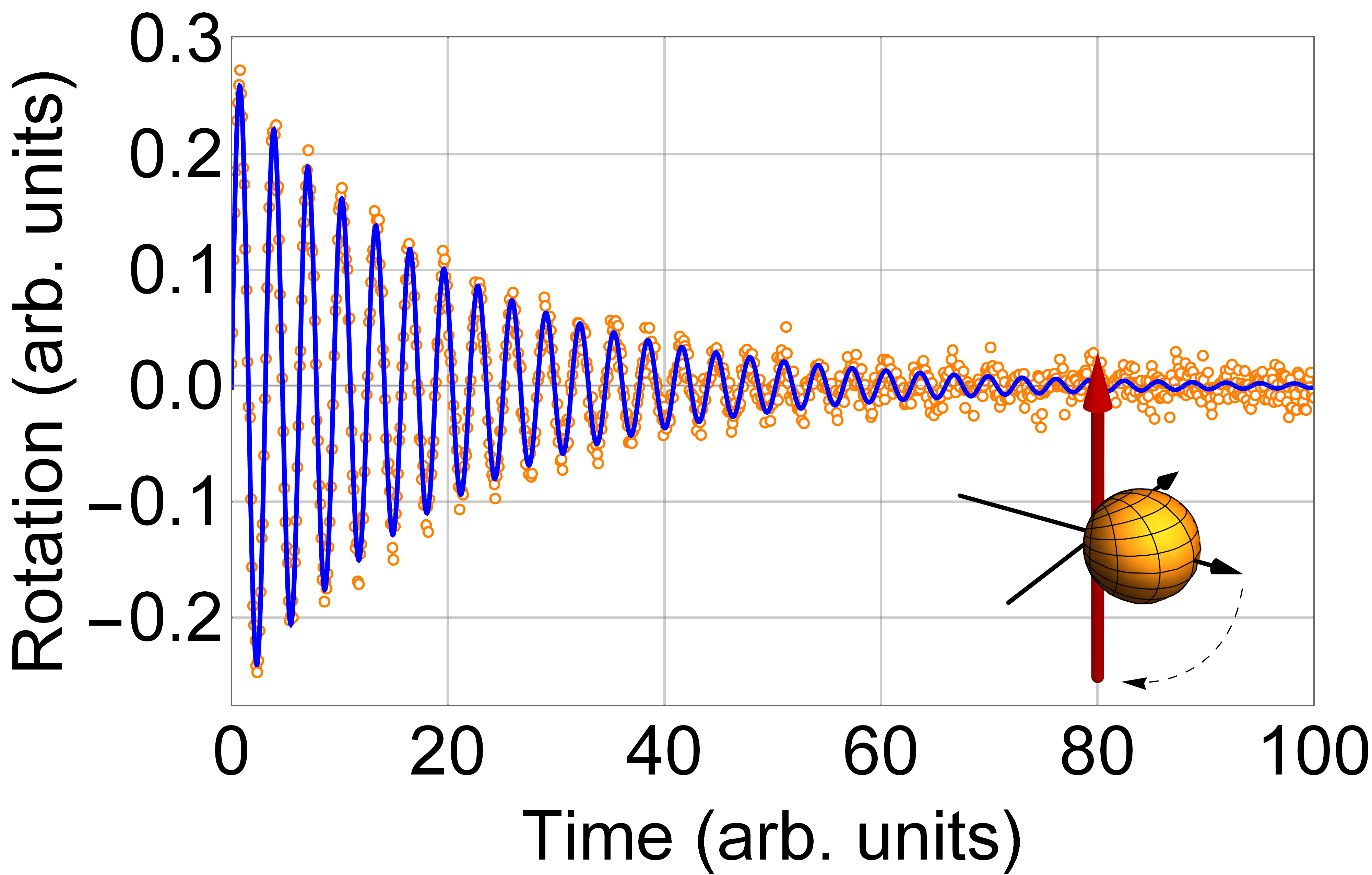} &   \includegraphics[width=0.3\textwidth]{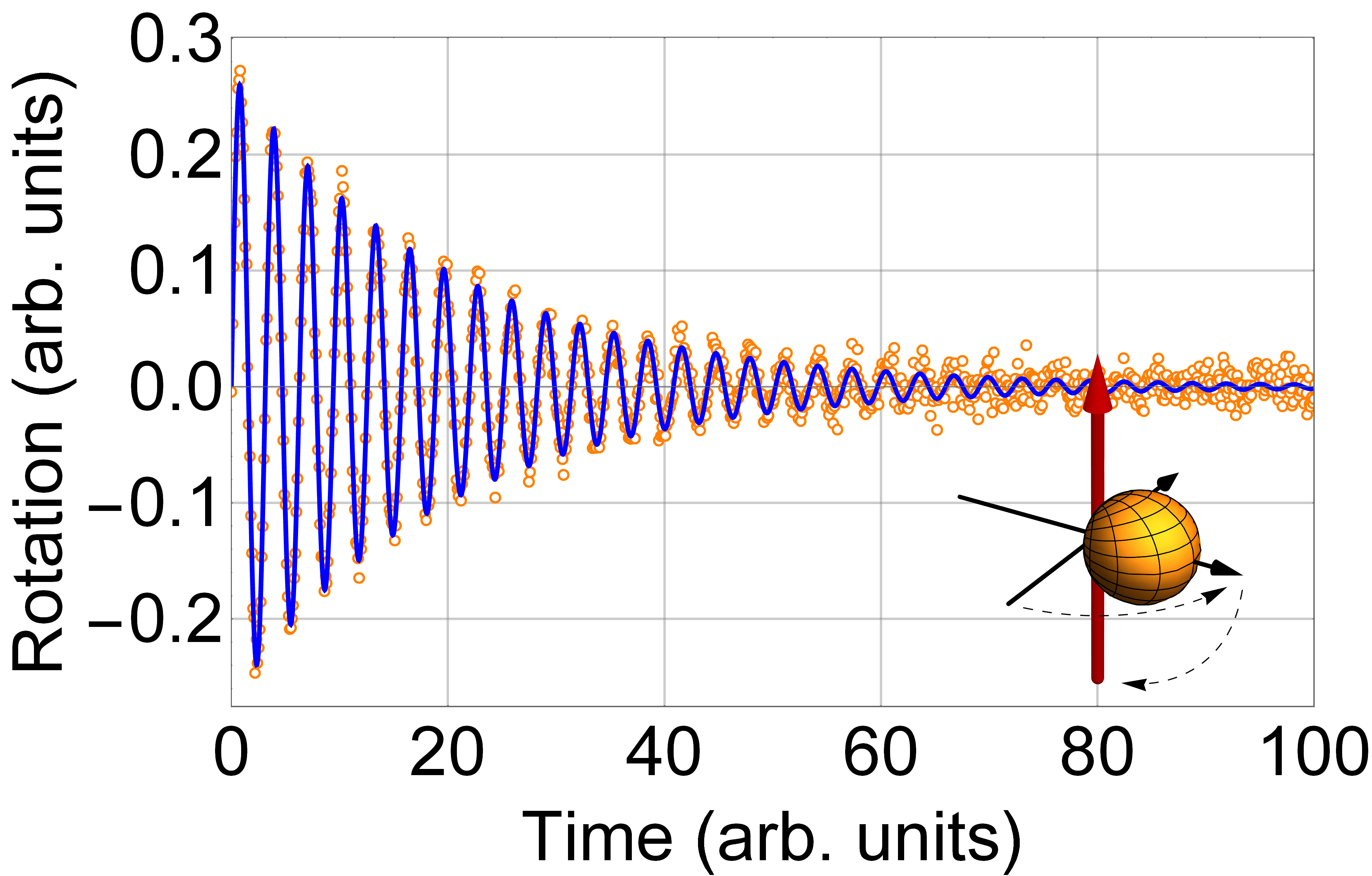} \\
        b) &  d) & f) \\[6pt]
    \end{tabular}
    \caption{Reconstruction of the density matrix corresponding to the stretch state. a) Absolute value of the density-matrix elements of the initial (orange/light grey) and reconstructed (blue/dark grey) states with  b) an absolute value of the differences between corresponding density-matrix elements. c)-f) Simulated rotation signals obtained with a set of four initial DC magnetic field pulses. }
    \label{fig:StretchState}
\end{figure}

\end{widetext}

The second nontrivial state, often encountered experimentally, is the fully oriented state (the stretch state).  This state is generated by circularly polarized light, which pumps atoms into the ground-state sublevel of a maximum or minimum magnetic number $m$ The corresponding density matrix is shown in Fig.~\ref{fig:StretchState}a). 
The analysis of the state using the AMPS shows that when the state is oriented along the magnetic field, there is no precession and hence no oscillating component of the signal. Then the only contribution to rotation signals comes from the population imbalance between the sublevels. Due to relaxation processes, the imbalance deteriorates over time.  As discussed in the main text, the population imbalance results in static rotation, one expects nonzero nonoscillating rotation, which decays over time.  Indeed, such behavior is observed in two of the data sets, corresponding to $\varphi=0$, $\theta=0$ and $\varphi=\pi/2$, $\theta=0$ pulse sets.  However, when the shape is oriented at an angle to the magnetic field, it provides an oscillating signal that is also visible in the simulations.  Based on the simulations [Fig.~\ref{fig:StretchState}c)-f)], the density matrix was reconstructed [blue bars in Fig.~\ref{fig:StretchState}a)].  The differences between the initial and the reconstructed matrix are shown in Fig.~\ref{fig:StretchState}b) and the fidelity of the reconstruction is equal to 0.999.

\bibliography{references}

\end{document}